\documentstyle[12pt,cite,epsf,epsfig,euscript,axodraw4j]{article}
\newcommand{\newc}{\newcommand}
\newc{\gev}{\,GeV}
\newc{\sgn}{\mr{sgn}\,}
\newc{\ra}{\rightarrow}
\newc{\rpv}{$\mathrm{\not\!R_p}$}
\newc{\met}{$\not\!\!E_T$}
\newc{\rp}{$\mathrm{R_p}$}
\newc{\real}{\mathcal{R}e}
\newc{\alsm}{{\displaystyle \sum_{\alpha=1,2}}}
\newc{\besm}{{\displaystyle \sum_{\beta=1,2}}}
\newc{\al}{\alpha}
\newc{\ga}{\gamma}
\newc{\de}{\delta}
\newc{\cw}{\cos\theta_w}
\newc{\ssw}{\sin^2\theta_w}
\newc{\ccw}{\cos^2\theta_w}
\newc{\cbe}{\cos\beta}
\newc{\sbe}{\sin\beta}
\newc{\sh}{\hat{s}}
\newc{\sa}{\sin\al}
\newc{\ca}{\cos\al}
\newc{\bv}{$\mathrm{\not\!B}$}
\newc{\lv}{$\mathrm{\not\!L}$}
\newc{\ie}{{\it i.e.\/}\ }
\newc{\lam}{\lambda}
\newc{\cht}{\tilde{\chi}}
\newc{\upt}{\tilde{u}}
\newc{\elt}{\tilde{\ell}}
\newc{\hgt}{\tilde{H}}
\newc{\nut}{\tilde{\nu}}
\newc{\dnt}{\tilde{d}}
\newc{\psb}{\bar{\psi}}
\newc{\rtt}{\sqrt{2}}
\newc{\mut}{\tilde{\mu}}
\newc{\mr}{\mathrm}
\newc{\bath}{\bar{\theta}}
\newc{\tht}{\theta}
\newc{\JC}{{\bf J}}
\newc{\lra}{\longrightarrow}
\newc{\eg}{{\it e.g.\,}}
\newc{\barr}{\begin{eqnarray}}
\newc{\earr}{\end{eqnarray}}
\newc{\beq}{\begin{equation}}
\newc{\eeq}{\end{equation}}
\newc{\me}{\mathcal{M}}
\newc{\dbm}{\partial_\mu}
\newc{\sgm}{\sigma_\mu}

\newcommand{\ep}{\mbox{$\epsilon$}}
\newc{\conv}{\otimes}
\newc{\pqqz}{P_{qq}^{(0)}}
\newc{\pqgz}{P_{qg}^{(0)}}
\newc{\pgqz}{P_{gq}^{(0)}}
\newc{\pggz}{P_{gg}^{(0)}}

\newcommand{\Dzero}{{\cal D}_0}
\newcommand{\Done}{{\cal D}_1}
\newcommand{\Dtwo}{{\cal D}_2}
\newcommand{\Dthree}{{\cal D}_3}
\newcommand{\deltaOmx}{\delta(1-x)}

\newcommand{\logTWqmuf}{\ln^2\left({m^2_{\phi}\over \mu_F^2}\right)}
\newcommand{\logqmuf}{\ln\left({m^2_{\phi}\over \mu_F^2}\right)}
\newcommand{\logTHOmx}{\ln^3\left(1-x\right)}
\newcommand{\logTWOmx}{\ln^2\left(1-x\right)}
\newcommand{\logONmx}{\ln\left(1-x\right)}

\newcommand{\logTWOpx}{\ln^2\left(1+x\right)}
\newcommand{\logONpx}{\ln\left(1+x\right)}
\newcommand{\logTHx}{\ln^3\left(x\right)}
\newcommand{\logTWx}{\ln^2\left(x\right)}
\newcommand{\logx}{\ln\left(x\right)}
\newcommand{\liTWONmx} {Li_2\left(1-x\right)}
\newcommand{\liTWmx} {Li_2\left(-x\right)}
\newcommand{\liTHOmx}{Li_3\left(1-x\right)}
\newcommand{\liTHmx}{Li_3\left(-x\right)}
\newcommand{\sONTWOmx}{S_{1,2}\left(1-x\right)}
\newcommand{\sONTWmx}{S_{1,2}\left(-x\right)}
\newcommand{\liTHpm}{Li_3\left({1-x \over 1+x}\right)}
\newcommand{\liTHmp}{Li_3\left(-{1-x \over 1+x}\right)}

\def\barr{\begin{array}}
\def\earr{\end{array}}
\def\be{\begin{equation}}
\def\ee{\end{equation}}
\def\ra{\rightarrow}

\catcode`@=11 
\def \gsim{\mathrel{\mathpalette\@versim>}}
\def \lsim{\mathrel{\mathpalette\@versim<}}
\def \@versim#1#2{\lower0.4ex\vbox{\baselineskip\z@skip\lineskip\z@skip
     \lineskiplimit\z@\ialign{$\m@th#1\hfil##\hfil$%
     \crcr#2\crcr\sim\crcr}}}
\catcode`@=12 
\def\gev{\: \rm GeV}

\def\etal{{\em et al.}}

\begin{document}
\setcounter{page}{0}
\renewcommand{\thefootnote}{\fnsymbol{footnote}}
\thispagestyle{empty}

\begin{titlepage}
\vspace{-2cm}
\begin{flushright}
RECAPP-HRI-2010-012\\[2ex]
{\large \tt hep-ph/yymmnnn}\\
\end{flushright}
\vspace{+2cm}

\begin{center}
{\Large{\bf NNLO QCD corrections to the resonant sneutrino/slepton production at 
Hadron Colliders}}\\
\vskip 0.6 cm
{{\bf Swapan Majhi}\footnote{swapan.majhi@saha.ac.in}
       {\rm and} {\bf Prakash Mathews}\footnote{prakash.mathews@saha.ac.in}\\
 Saha Institute of Nuclear Physics, 1/AF Bidhannagar, \\Kolkata 700064,
        India.}
\vskip 0.2 cm
{ {\bf V Ravindran}\footnote{ravindra@hri.res.in}\\
Regional Centre for Accelerator based Particle Physics,\\
Harish-Chandra Research Institute, Chhatnag Road, Jhusi,\\ 
Allahabad 211 019, India.
}
\vskip 0.2 cm
\end{center}
\setcounter{footnote}{0}
\begin{abstract}\noindent
We present a complete next to next to leading order QCD corrections to
the resonant sneutrino and charged slepton production cross sections 
at hadronic colliders such as the Tevatron and the Large Hadron Collider  
within the context of $R$-parity violating
supersymmetric model.  We have demonstrated the role of
these corrections in reducing uncertainties resulting from 
renormalisation and factorisation scales and thereby making our predictions reliable. 
We have incorporated soft gluon effects at $N^3LO$ level in order to study the stability 
of our results under perturbation.  The NNLO corrections are found to be
large and significant.  
The results obtained in this article are also applicable to 
resonance production of any color-neutral scalar. 
\end{abstract}
\end{titlepage}

\setcounter{footnote}{0}
\renewcommand{\thefootnote}{\arabic{footnote}}

\setcounter{page}{1}
\pagestyle{plain}
\advance \parskip by 10pt
\section{Introduction}
The Standard Model (SM) of particle physics is undoubtedly a very successful model
even though its scalar sector remains in an unsatisfactory state of affairs.
This is because of the fact that the Higgs boson in the scalar sector 
which is responsible for symmetry breaking mechanism is still missing.
There are also other issues such as gauge hierarchy problem, 
dark matter, baryogenesis, gauge unification etc., that are still not
fully understood within the framework of the SM, pointing to
physics beyond the standard model.
In the past, the above mentioned issues were being addressed 
by going beyond the SM.  Two of the most attractive classes of such models
contain features incorporating supersymmetry~\cite{SUSY} and/or
grand unification~\cite{GUT} (scenarios with a low
intermediate scale~\cite{framp}).  These models contain
large number of new particles including new scalars. 
The couplings of the first
generation SM fermions to these new scalars need not be suppressed. 
This provides us a new platform to study their potential new physics signals.

The electroweak gauge invariance ensures both
baryon ($B$) and lepton ($L$) number conservations within the SM,
but this is not the case for supersymmetry (SUSY). The most general 
superpotential respecting the gauge symmetry of the SM contains bilinear 
and trilinear terms which do not respect $B$ or 
$L$ conservations. 
A discrete symmetry called {\it R-}parity can be used to  
forbid such terms.  The
corresponding conserved quantum number is given by $\it R_p \equiv (-1)^{3B+L+ 2 S}$, 
where $S$ is the spin of the particle.
In the minimal supersymmetric standard model (MSSM), this symmetry has
been originally imposed to suppress the rapid proton decay, 
it also guarantees stability of lightest supersymmetric particle (LSP) and 
hence a natural candidate for cold dark matter.

The introduction of this symmetry is {\it not} the only option to suppress proton decay,  
there exists several alternatives through introduction of other symmetries.
Hence, it is of phenomenological interest to consider possible violations
of R-parity and study their experimental consequences. 

The possible $R$-parity violating (\rpv)
terms in the superpotential can be parametrised as
\be
    {\cal W}_{\not R_p} = \mu_i L_i H_2
                        + \lambda_{ijk} L_i L_j E^c_k
                        +  \lambda'_{ijk} L_i Q_j D^c_k
                        +  \lambda''_{ijk} U^c_i D^c_j D^c_k \ ,
      \label{eq:superpot}
\ee
where $L_i$ and $Q_i $ are the $SU(2)$-doublet lepton and quark
superfields, $E^c_i, U^c_i, D^c_i$
the singlet superfields and $H_i$ the Higgs superfields.
Note that $\lambda_{ijk}$ is antisymmetric
under the interchange of the first two indices and $\lambda''_{ijk}$
is antisymmetric under the interchange of the last two.
The first three terms in eqn.(\ref{eq:superpot}) violate
$L$ and the last term violates $B$ conservation.
We need to have at least one of the two sets of couplings to be vanishingly small
in order to satisfy the constraints coming from the non--observance of proton decay.
In the following, we assume that $B$ is a good symmetry
of the theory, or in other words all of $\lambda''_{ijk}$ are zero.
This can also suppress all dimension six operators leading
to proton decay (see~\cite{Ibanez:1992pr}) along with the
dimension five ones. 
In \cite{hall-suzuki,Ibanez:1992pr},
this scenario has been motivated within certain theoretical
frameworks and makes simpler, 
problem of preservation of GUT--scale baryon asymmetry~\cite{baryo}.
The presence of the other \rpv\  terms 
can affect the baryon asymmetry of the universe.  Fortunately
these bounds are highly model-dependent and hence 
we evade them (see~\cite{dreiner-ross}). 


Each interaction in eqn.(\ref{eq:superpot}) has its unique 
experimental signatures through low-energy phenomenology and/or in resonance
production at colliders.  
Large number of bounds on these coupling result from the studies
of low and intermediate energy processes, namely
rare decays with lepton and/or hadron flavor violation.
The neutral and charged current universality in lepton and quark sectors,
high precision measurements in anomalous magnetic and electric dipole
moments and observables with CP violation provide  
stringent constraints on these couplings at low energies.
Terms involving $\lambda_{ijk}$ lead to resonant
sneutrino production in
$e^+ e^-$ collider~\cite{resonance_ee,Barger:1989rk} and 
those involving $\lambda''_{ijk}$ lead to
resonant squark production in hadron--hadron collisions
  \cite{slepton_squark_reso,squark_reso}.
Similarly, the $\lambda'_{ijk}$ terms can contribute to both
resonant squark production at an $e^\pm p$ colliders~\cite{HERA_expl}
as well as to resonant charged slepton and sneutrino production at
hadron colliders~\cite{slepton_squark_reso,slepton_reso}.
A simplified search strategy at colliders could be to assume the existence
of one dominant R-parity violating coupling at a time.  In the past
most of the studies rely on this assumption.
Since the absence of tree-level flavor changing neutral current processes
lead to severe constraints on the simultaneous presence
of more than one $\lambda'_{ijk}$~\cite{fcnc}, we shall 
restrict ourselves to only one non-zero $\lambda'_{ijk}$. 
In the rest of the paper, we will concentrate only on the $\lambda'_{ijk}$ couplings 
that can affect the resonant  production.  That part of the Lagrangian can be 
written in terms of the component fields as
\be
\barr{rcl}
   {\cal L}_{\lambda'} & = & \lambda'_{ijk}
            \left[ \overline{d_{kR} } \nu_{iL} \tilde{d}_{jL} +
                   \overline{d_{kR} } d_{jL} \tilde{\nu}_{iL} +
                   \overline{(\nu_{iL})^c } d_{jL} \tilde{d}^\ast_{kR}
            \right. \\[1.5ex]
& & \left. \hspace*{1.5em}
                   - \overline{d_{kR} } \ell_{iL} \tilde{u}_{jL} -
                   \overline{d_{kR} } u_{jL} \tilde{\ell}_{iL} -
                   \overline{(\ell_{iL})^c } u_{jL} \tilde{d}^\ast_{kR}
            \right] + {\rm h.c.}
\earr
      \label{lambda-pr}
\ee
The squarks in supersymmetric theories behave as leptoquarks and
the charged sleptons/sneutrinos behave as
charged/neutral Higgses of multi-Higgs-doublet
scenario.   Hence, one would expect non-zero values for these couplings 
will have important phenomenological consequences.

Quantum Chromodynamics (QCD) plays an important role in hadron colliders
as the underlying scattering processes involve quarks (anti-quarks) and    
gluons.  The leading order (LO) scattering processes involving quarks (anti-quarks) and
gluons often predict results that are sensitive to large theoretical uncertainties
through non-perturbative parton densities and missing higher order perturbative
radiative corrections.  In the next sections, we will elaborate on 
the sources of these uncertainties and
provide systematic methods to reduce them.
In \cite{swapan},  first results on the next-to-leading order (NLO) 
QCD corrections to sneutrino and charged slepton productions at hadron colliders 
were reported.   These results were later confirmed and further NLO 
SUSY-QCD corrections were systematically incorporated in the analysis by 
the authors of \cite{dreiner}. 
It was found that
the NLO QCD effects were quite large $\sim 10\% - 40\%$ at both Tevatron as well as 
LHC and hence they were used by both CDF\cite{CDF_col} and 
D0\cite{D0_col} collaboration to
analyse their data (Run-I as well as Run-II data). 
In their analysis to set bound on these R-parity violating couplings, 
K-factor for
SM background 
\cite{} 
was considered at the next to next to leading order (NNLO) level while for the R-parity violating effects only NLO
K-factor was used.     
Therefore, it is desirable to compute the K-factors 
for the resonant sneutrino and/or charged slepton productions at NNLO in QCD. 
These results will quantitatively improve the analysis
based on high statistics data available in the ongoing and future
experiments.  From the theoretical point of view, higher order 
radiative corrections provide a test of the convergence of the perturbation theory and
hence the reliable comparison of data with the theory predictions is possible. 
We have also studied the effects of soft gluons which often dominate 
at hadronic collisions.  This also opens up the possibility of resumming them 
to all orders in perturbation theory through suitable framework.
The fixed order perturbative results most often suffer from large uncertainties 
due to the presence of renormalisation and factorisation scales. 
They get reduced as we include more and more terms
in the perturbative expansion thanks to renormalisation group
invariance.  It is also important to estimate the uncertainties coming from the
choice of parton density sets available in the literature.
Due to these reasons, we think that it is worthwhile to perform the next-next-to-leading order 
(NNLO) calculation to the aforementioned processes. In our previous paper\cite{swapan}, we
used fixed \rpv coupling ($\lambda'=0.01$) based on the assumption that 
its scale dependence is very weak, in addition the {\em value} of the coupling $\lambda'$
was immaterial and only served to set an overall
normalisation for the cross-section.  In the present paper we have
systematically included its scale dependence through the renormalisation group equations 
and we will discuss the impact of it in the next sections. 

We briefly review the constraints on $\lambda'$ from low-energy phenomenology.
Non-zero $\lambda'$s can lead to additional
four-fermion operators that may contribute to meson decays, neutral
meson mixing, some of which may be forbidden otherwise. 
In Table~\ref{tab:coup_limits}, we list the currently known bounds
on several of these couplings\footnote{A more complete list can be found
        in refs.\protect\cite{bounds}.}.
The strongest bound is on $\lambda'_{111}$
and is derived from non-observation of neutrinoless
double beta decay $(a)$~\cite{bb0nu}. The others are much weaker and
are derived from
        ($b$) upper bound on the mass of the
                $\nu_e$~\cite{hall-suzuki,belesev,gb_dc,anjan};
        data on ($c$) charged-current universality~\cite{Barger:1989rk};
                ($d$) atomic parity violation~\cite{APV};
                ($e$) $\tau \ra \pi \nu_\tau$ and
                      $D \ra K l \nu$~\cite{gb_dc};
        and     ($f$) $D^0$-$\overline{D^0}$ mixing~\cite{fcnc}.
\begin{table}[htb]
\begin{center}
\bigskip
\begin{tabular}{||c|c|| c|c|| c|c||}
\hline
$\{ijk\}$ & Existing bounds & 
$\{ijk\}$ & Existing bounds & 
$\{ijk\}$ & Existing bounds \\[1ex]
\hline
111 & 0.001$^{\:a)}$ &
     211 & 0.09$^{\:c)}$ &
        311  & 0.10$^{\:e)}$ \\
112 & 0.02$^{\:c)}$ &
     212 & 0.09$^{\:c)}$ &
        312  & 0.10$^{\:e)}$ \\
121 & 0.035$^{\:d)}$ &
     221 & 0.18$^{\:e)}$ &
        321  & 0.20$^{\:f)}$ \\
122 & 0.02$^{\:b)}$ &
     222 & 0.18$^{\:e)}$ &
        322  & 0.20$^{\:f)}$ \\
\hline
\end{tabular}
\caption[] {The  upper bounds on the $\lambda'$--type \rpv\ couplings
of interest for a common sfermion mass $\tilde{m} = 100$ GeV. 
The superscripts refer to the specific experiments 
leading to the constraints and as described in the text.
}
\label{tab:coup_limits}
\end{center}
\end{table}

Since these bounds are derived from effective four-fermion operators,
they typically scale like the mass of the exchanged
sfermion\footnote{Of those listed in Table~\protect\ref{tab:coup_limits},
        the only exceptions to this rule are the bounds for
        $\lambda'_{111}$ and
        $\lambda'_{122}$~\protect\cite{bb0nu,hall-suzuki,belesev,gb_dc}.}.
These bounds are actually
applicable only to particular combinations of couplings and masses and
reduce to those in the table only under the assumption of only one coupling
being non-zero. In addition, in meson decays, most often it is the squark
that is exchanged; hence charged sleptons/sneutrinos could very well be much lighter
without contradicting the bounds.

In the next section, we describe the computation of 
resonant production of scalar/pseudo scalar 
to NNLO in perturbative QCD.  
The results are obtained in the $\overline{MS}$ scheme. We then proceed to
study the impact of our results on their production cross sections for both
Tevatron and LHC energies.  Finally we summarise our findings in the conclusion.         
Our analytical results that go into the numerical code are presented in the
Appendix.
%
\section{Computation of partonic coefficient functions to order $\alpha_s^2$}
        \label{sec:LO}

In this section, we describe in detail, the computation of second order ($\alpha_s^2$) 
QCD radiative corrections to resonant production, in hadron colliders, 
of a neutral scalar particle $\phi(\xi)$ which couple to fermionic fields 
$\psi(\xi)$ through Yukawa interaction given by the action:
\begin{eqnarray}
S=\int d^4 \xi~ {\cal L}_{int}(\xi) = \lambda' \int d^4 \xi ~ \phi (\xi)~ \overline \psi(\xi) \psi(\xi)
\end{eqnarray}
where $\lambda'$ is the coupling strength of the interaction.
We present our results in such a way that 
they can be used for a similar study on charged scalar production as well and hence
are applicable to a detailed study on resonant production of sneutrinos and charged sleptons which
is the main goal of this present work.  The inclusive hadronic cross section for the reaction
\begin{eqnarray}
\label{eqn2.10}
H_1(P_1)+H_2(P_2)\rightarrow \phi (p_5)+X\,,
\end{eqnarray}
 is given by
\begin{eqnarray}
\label{eqn2.11}
&&\sigma_{\rm tot}^{\phi}={\pi {\lambda'}^{2}(\mu_R^2) \over 12 S} \sum_{a,b=q,\bar q,g}\,
\int_\tau^1 {dx_1 \over x_1}\, \int_{\tau/x_1}^1{dx_2\over x_2}\,f_a(x_1,\mu_F^2)\,f_b(x_2,\mu_F^2)\,
\Delta_{ab}\left ( \frac{\tau}{x_1\,x_2},m^2_{\phi},\mu_F^2,\mu_R^2 \right ) 
\nonumber\\[2ex]
&&\mbox{with}\quad \tau=\frac{m^2_{\phi}}{S} \quad\,,\quad S=(P_1+P_2)^2\quad
\,,\quad p_5^2=m^2_{\phi}\,,
\label{sigphi}
\end{eqnarray}
where $H_1$ and $H_2$ denote the incoming hadrons and $X$ represents an
inclusive hadronic state.
The parton densities denoted by
$f_c(x_i,\mu_F^2)$ ($c=q,\bar q,g$)
depend on the scaling variables $x_i$ ($i=1,2$) through $p_i=x_i P_i$ and 
the mass factorization scale $\mu_F$.  Here $p_i$ ($i=1,2$) are
the momenta of incoming partons namely quarks, anti-quarks and gluons. 
The coupling constant $\lambda'$ gets renormalised at the renormalisation scale
$\mu_R$ due to ultraviolet singularities present in the theory.
The factorisation scale is introduced on the right hand side of the
above equation to separate long distant dynamics from the perturbatively 
calculable short distant partonic coefficient 
functions $\Delta_{ab}$.  $\Delta_{ab}$ depends on both $\mu_R$ and $\mu_F$ 
in such a way that the entire scale dependence goes away to all orders
in perturbation theory when convoluted with appropriate parton densities.
This is due to the fact that the observable on the left hand side of the above equation 
is renormalisation group (RG) invariant with respect to both the scales.  This implies
\begin{eqnarray}
&&\mu^2{d \sigma_{\rm tot}^{\phi} \over d \mu^2} =0, \quad \quad \mu=\mu_F,\mu_R \, ,
\\[2ex]
&&\mu_R^ 2{d 
\over d \mu_R^2} \Big[{\lambda'}^2(\mu_R^2) \Delta_{ab}\left (x,m^2_{\phi},\mu_F^2,\mu_R^2\right ) 
\Big]=0\, .
\label{RGE}
\end{eqnarray}
The partonic coefficient functions that appear in eqn.(\ref{sigphi})
are computable in perturbative QCD in terms of strong coupling constant $g_s$.
The ultraviolet singularities present in the theory are regularised in dimensional
regularisation and are removed in ${\overline {MS}}$ scheme, introducing 
the renormalisation scale $\mu_R$ at every order in perturbative expansion.  
In addition, the Yukawa coupling $\lambda'$ also gets renormalised due to 
strong interaction dynamics.  Hence, for our computation, we require only 
two renormalisation constants 
to obtain UV finite partonic coefficient functions, $\Delta_{ab}$.  
These constants are denoted by
$Z(\mu_R)$ and $Z_{\lambda'}(\mu_R)$, where the former renormalises the strong coupling constant $g_s$ 
and the later Yukawa coupling $\lambda'$. 

We define the bare strong coupling constant by 
$\hat a_s={\hat g^2_s/ 16 \pi^2} $,
$\hat g_s$ being the dimensionless strong coupling constant in
$n=4+\ep$, with $n$ being the number of space time dimensions.
The bare coupling constant $\hat a_s$ is related to the renormalised one, 
$a_s(\mu_R^2)$ by the following relation:
\begin{eqnarray}
S_{\ep} \hat a_s = Z(\hat a_s,\mu^2,\mu_R^2,\ep) a_s(\mu_R^2) \left(\mu^2 \over \mu_R^2\right)^{\ep \over 2}\,.
\label{renas}
\end{eqnarray}
The scale $\mu$ comes from
the dimensional regularisation in order to make the bare coupling constant $\hat g_s$
dimensionless in $n$ dimensions.
$S_{\ep}$ is the spherical factor characteristic of $n$-dimensional regularisation.

The renormalisation constant that relates the bare coupling constant
$\hat a_s$ to the renormalised one $a_s(\mu_R^2)$ through
the eqn.(\ref{renas}) is given by
\begin{eqnarray}
Z(\hat a_s,\mu^2,\mu_R^2,\ep)= 1&+& \hat a_s\Bigg({\mu_R^2 \over \mu^2}\Bigg)^{{1 \over 2}\ep} S_{\ep}
          \Bigg[{2 \beta_0 \over \ep}\Bigg]
           + \hat a_s^2 \Bigg({\mu_R^2 \over \mu^2}\Bigg)^{\ep} S^2_{\ep} \Bigg[
                  {\beta_1 \over \ep} \Bigg]
\nonumber\\[2ex]
           &+& \hat a_s^3\Bigg({\mu_R^2 \over \mu^2}\Bigg)^{{3 \over 2}\ep} S^3_{\ep} \Bigg[ 
                   -{4 \beta_0~ \beta_1 \over 3 \ep^2}
                   +{2 \beta_2 \over 3 \ep}\Bigg] +{\cal O}(\hat a_s^4) \,.
\end{eqnarray}
The coefficients $\beta_i$ for $i=1,...,4$ 
can be found in \cite{4loop_beta_func}
for $SU(N)$ QCD 
expressed in terms of the color factors:  
\begin{eqnarray}
C_A=N,\quad \quad \quad C_F={N^2-1 \over 2 N} , \quad \quad \quad
T_F={1 \over 2}\,.
\end{eqnarray}
and $n_f$ is the number of active flavors.
Similarly, for the Yukawa coupling, we have 
\begin{eqnarray}
S_{\ep}{\hat \lambda'} = Z_{\lambda'}(\hat a_s,\mu^2,\mu_R^2,\ep) {\lambda'}(\mu_R^2)
 \left(\mu^2 \over \mu_R^2\right)^{\ep \over 2} \,,
\label{ren_lamb}
\end{eqnarray}
where
\begin{eqnarray}
Z_{\lambda'}(\hat a_s,\mu^2,\mu_R^2,\ep)& =&1
    + \hat a_s \left({\mu_R^2 \over \mu^2}\right)^{{\ep \over 2}} S_{\ep}
       \Bigg[ {1 \over \ep}   \Bigg( 2~ \gamma_0 \Bigg)\Bigg]
   +\hat a_s^2 \left({\mu_R^2 \over \mu^2}\right)^{{\ep }} S_{\ep}^2
       \Bigg[ {1 \over \ep^2}   \Bigg( 2~ \Big(\gamma_0\Big)^2
           - 2~ \beta_0~ \gamma_0 \Bigg)
\nonumber\\[2ex]
&&       + {1 \over \ep}   \Bigg( \gamma_1 \Bigg)\Bigg]
    +\hat a_s^3 \left({\mu_R^2 \over \mu^2}\right)^{3{ \ep \over 2}} S_{\ep}^3
       \Bigg[ {1\over \ep^3}   \Bigg( {4\over 3}~ \Big(\gamma_0\Big)^3
        - 4~ \beta_0~ \Big(\gamma_0\Big)^2
          + {8\over 3}~ \beta_0^2~ \gamma_0 \Bigg)
\nonumber\\[2ex]
&&       + {1 \over \ep^2}   \Bigg( 2 ~\gamma_0 ~\gamma_1
       - {2\over 3}~ \beta_1~ \gamma_0
       - {8\over 3}~ \beta_0 ~\gamma_1 \Bigg)
       + {1 \over \ep}   \Bigg( {2\over 3} ~\gamma_2 \Bigg)\Bigg]
       +{\cal O}(\hat a_s^4)\,.
\label{Z_lamb}
\end{eqnarray}
The anomalous dimensions $\gamma_i$ for $i=1,...,4$ can be obtained from
the quark mass anomalous dimensions given in 
\cite{vanRitbergen:1997va}.  
While the ${\cal O}(\hat a_s^3)$ terms in both $Z$ and $Z_{\lambda'}$ 
do not contribute to partonic sub processes
computed to order $a_s^2$,  they determine the scale evolution of 
both $a_s$ and $\lambda'$ to NNLO through renormalisation group equations:
\begin{eqnarray}
\mu_R^2 {d \over d\mu_R^2} \ln a_s(\mu_R^2) &=&
-\sum_{i=1}^\infty a^i_s(\mu_R^2)~ \beta_{i-1} \,,
\nonumber\\
\mu_R^2 {d \over d\mu_R^2} \ln \lambda'(\mu_R^2)&=&
-\sum_{i=1}^\infty a^i_s(\mu_R^2)~ \gamma_{i-1}\,.
\label{REASLAM}
\end{eqnarray}
and constitute dominant soft gluon contribution to order $\alpha_s^3$.
The perturbatively calculable $\Delta_{ab}$ can be expanded in powers of
strong coupling constant $a_s(\mu_R^2)$ as
\begin{eqnarray}
\Delta_{ab}\left (x,m^2_{\phi},\mu_F^2,\mu_R^2\right )=\sum_{i=0}^\infty a_s^i(\mu_R^2)
\Delta_{ab}^{(i)}\left (x,m^2_{\phi},\mu_F^2,\mu_R^2\right )
\nonumber
\end{eqnarray}
$\Delta_{ab}$ gets contributions from various partonic reactions.

Having studied the UV renormalisation constants relevant for our computation, we now
list out various partonic sub processes that will contribute to NNLO order ($\alpha_s^2$).
The leading order partonic reaction (fig.(\ref{fig1})) is given by  
\begin{eqnarray}
q_i +\overline q_j 
\rightarrow \phi 
\end{eqnarray}
and its contribution is found to be proportional to $\delta(1-x)$ where
$x=m_\phi^2/s$ with partonic center of mass energy $s=(p_1+p_2)^2$. 
At NLO (fig.(\ref{fig2},\ref{fig3},\ref{fig4})), we have    
\begin{eqnarray}
q_i +\overline q_j &\rightarrow& \phi+ {\rm one ~~ \rm loop},
\nonumber\\[2ex]
q_i +\overline q_j &\rightarrow& \phi+ g, 
\nonumber\\[2ex]  
q_i(\overline q_i) +g &\rightarrow&\phi+ q_j(\overline q_j) .
\label{qqbnloloop}
\end{eqnarray}
The subscripts $i,j$ in the quark (anti-quark) fields denote the flavor indices.
If  $i=j$, then $\phi$ will be neutral scalar otherwise it will denote a charged scalar.   

At NNLO, there are several new channels
open up and we discuss them one by one: 
\begin{itemize}
\item
processes with quark anti-quark pair in the initial state with no gluon (fig.(\ref{fig5})), one gluon (fig.(\ref{fig6})) 
and two gluons (fig.(\ref{fig7})) in the final state along with $\phi$:
\begin{eqnarray}
q_i +\overline q_j &\rightarrow& \phi + {\rm two ~~ \rm loop} \,,
\nonumber\\[2ex]
q_i +\overline q_j &\rightarrow& \phi +g+ {\rm one ~~ \rm loop} \,,
\nonumber\\[2ex]
q_i +\overline q_j &\rightarrow& \phi +g + g \,.
\label{qqbnnloloop}
\end{eqnarray}
\item
processes (fig.(\ref{fig8})) with quark anti-quark pair in the final states along with $\phi$:
\begin{eqnarray}
q_i +\overline q_j \rightarrow \phi + q_k + \overline q_l \,,
\label{reactionab}
\end{eqnarray}
and quark (anti-quark) quark (anti-quark) pair
in the final state with $\phi$ (fig.(\ref{fig9})):  
\begin{eqnarray}
q_i +q_j \rightarrow \phi + q_k +  q_l \,,
\nonumber\\[2ex]
\overline q_i +\overline q_j \rightarrow  \phi + \overline q_k +  \overline q_l \,.
\label{reactioncd}
\end{eqnarray}
The contributions coming from the last two reactions (eqn.(\ref{reactioncd})) 
will have two possibilities: the final states with identical
quarks (anti-quarks) or non-identical quarks (anti-quarks).  
For identical quarks (anti-quarks) in the final state (fig.(\ref{fig10})),
the number of $t$ channel processes doubles as we need to
include processes with final state quarks (anti-quarks) interchanged.  
We also need to appropriately multiply the statistical factor $1/2$.
\item
processes with quark (anti-quark) and gluon in the initial states 
\begin{eqnarray}
q_i (\overline q_i) + g &\rightarrow& \phi+q_j(\overline q_j)  + 
{\rm one} \quad {\rm loop} \,,
\nonumber\\[2ex]
q_i (\overline q_i) + g &\rightarrow& \phi+q_j(\overline q_j) + g \,.
\end{eqnarray}
\item
processes with pair of gluons in the initial state 
\begin{eqnarray}
g+ g \rightarrow \phi+ q_i + \overline q_j \,.
\end{eqnarray}
\end{itemize}

The calculation of various contributions from the partonic reactions 
involves careful handling of divergences that result from
one\cite{1loop} and two loop\cite{2loop} integrations in the virtual
 processes  
and two and three body phase space integrations in the real emission processes.
The loop integrals often give ultraviolet, soft and collinear
divergences.  But the phase space integrals give only soft and collinear 
singularities.  Soft divergences arise when the  
momenta of the gluons become zero while the collinear diverges
arise due to the presence of massless partons.   We have regulated all the integrals in
dimensional regularisation with space time dimension $n=4+\ep$.  
The singularities manifest themselves as poles in $\ep$.

We have reduced all the one loop tensorial integrals
to scalar
integrals using the method of Passarino-Veltman \cite{Pas_Velt} in $4+\ep$
 dimensions
and evaluated resultant scalar integrals exactly.  
The form factor that contributes to reactions in eqn (\ref{qqbnnloloop})
is obtained using the dispersion
technique \cite{Cutkosky:1960sp}
and is presented in the appendix.
Two and three body phase space integrals are done
by choosing appropriate Lorentz frames\cite{2.3.Body_phase_space}.
Since we integrate over the total phase space the integrals are Lorentz
invariant and therefore frame independent. 
Several routines are made using the algebraic manipulation program FORM\cite{form} 
in order to perform tensorial reduction of one loop integrals and 
two and three body phase space integrals.

The UV singularities go away after performing renormalisation through
the constants $Z$ and $Z_{\lambda'}$.  
The soft singularities cancel among virtual and real emission processes\cite{BN}
at every order in perturbation theory.  The remaining 
collinear singularities are renormalised systematically using
mass factorisation\cite{KLN} as follows.  
For more details on the computation of NNLO QCD corrections to process of
the kind considered here can be found in \cite{nnlody}. 
Let us denote the resulting UV and soft 
finite partonic cross sections as    
\begin{eqnarray}
\hat \Delta_{ab}\left (x,m^2_{\phi},\mu_R^2 \right )=
\frac{s}{{\lambda'}^{2}(\mu_R^2)}\,
\hat \sigma_{ab,{\phi}}\left(x,m^2_{\phi},\mu_R^2\right)\,,
\nonumber
\end{eqnarray}
where
\begin{eqnarray}
\label{eqn3.1}
\hat \sigma_{ab,{\phi}}\left(x,m^2_{\phi},\mu_R^2\right)&=&K_{ab}\,\frac{\pi}{s}\,Z_{\lambda'}^2\,
\int d^np_5\,\delta^+(p_5^2-m^2)
\sum_{m=3}^{\infty}\,\prod_{i=3,i\not =5}^m\int\frac{d^np_i}{(2\pi)^{n-1}}\,
\delta^+(p_i^2)
\nonumber\\[2ex]
&&\qquad\qquad\times
\delta^{(n)}(\sum_{j=1}^m\,p_j)\, |M_{ab\rightarrow X~{\phi}}|^2\,,
\nonumber\\[2ex]
M_{ab\rightarrow X~{\phi}}&=&\langle p_1,p_2|\hat O(0)| X,p_5\rangle
\quad \mbox{with} \quad | X,p_5\rangle=|p_3,p_4,p_6 \cdots p_m,p_5\rangle\,,
\nonumber\\[2ex]
\end{eqnarray}
$\hat O(0)$ is the interaction operator responsible for the reactions
with scalars and $K_{ab}$ represents the spin and colour average over the initial states.
Mass factorisation allows us to express the collinear singular 
partonic cross section $\hat \Delta_{ab}$ in terms of pair of singular
transition functions $\Gamma_{cd}(x,\mu_F^2,\ep)$, namely Altarelli-Parisi kernels and finite 
partonic coefficient function $\Delta_{ab}$:   
\begin{eqnarray}
\label{eqn2.11}
\hat \Delta_{ab}\left (x,m^2_{\phi},\mu_R^2\right )&=&  \sum_{c,d=q,\bar q,g}\,
\int_x^1 {dx_1 \over x_1}\, \int_{x/x_1}^1{dx_2\over x_2}\,\Gamma_{ca}(x_1,\mu_F^2,\ep)\,\Gamma_{db}(x_2,\mu_F^2,\ep)\,
\nonumber\\[2ex] && \quad \quad \quad \qquad\qquad\times
\Delta_{cd}\left ( \frac{x}{x_1\,x_2},m^2_{\phi},\mu_F^2,\mu_R^2 \right )\,.
\label{massfact}
\end{eqnarray}
The transition functions are perturbatively calculable in powers of $a_s(\mu_F^2)$: 
\begin{eqnarray}
\Gamma_{ab}(x,\mu_F^2,\ep) = \sum_{i=0}^\infty a_s^{i}(\mu_F^2) \Gamma^{(i)}_{ab} (x,\ep)\,.
\end{eqnarray}
In $\overline{MS}$ mass factorisation scheme, they are found to be (suppressing the
arguments $x$ and $\ep$)
\begin{eqnarray}
\Gamma_{ab}^{(0)} &=& \delta_{ab} \delta(1-x)\,,
\\[2ex]
\Gamma_{ab}^{(1)} &=& -{1 \over \epsilon} P^{(0)}_{ab}\,,
\end{eqnarray}
and
\begin{eqnarray}
\Gamma_{ab}^{(2)}={1 \over 2 \epsilon^2}
\sum_{c}\Bigg( P_{ac}^{(0)}\otimes P_{cb}^{(0)} + 2 \beta_0 P_{ab}^{(0)}
\Bigg)
+{1 \over 2 \epsilon} P_{ab}^{(1)} \,.
\end{eqnarray}
More explicitly we have:
\begin{eqnarray}
\Gamma_{q q}^{(2),NS} &=&\Gamma_{\bar q \bar q}^{(2),NS} =
\bigg[{1 \over 2 \epsilon^2}\Bigg( P^{(0)}_{qq}\otimes P^{(0)}_{qq} 
+2 \beta_0 P^{(0)}_{qq}(x)\Bigg)+
 {1 \over 2 \epsilon} P_{qq}^{(1),NS}\Bigg]
\nonumber\\[2ex]
\Gamma_{q \bar q}^{(2),NS} &=&\Gamma_{\bar q q}^{(2),NS} =
 {1 \over 2 \epsilon} P_{qq}^{(1),-}
\nonumber\\[2ex]
\Gamma_{q q}^{(2),S/-} &=&\Gamma_{\bar q \bar q}^{(2),S/-} =  
\Gamma_{q \bar q}^{(2),S/-} =\Gamma_{\bar q q}^{(2),S/-} =  
\Bigg[{1 \over 2 \epsilon^2} P^{(0)}_{qg}\otimes P^{(0)}_{gq} + {1 \over 2 \epsilon}P_{qq}^{(1),S/-}
\Bigg]
\nonumber\\[2ex]
\Gamma_{q g}^{(2)} &=&\Gamma_{\overline q g}^{(2)} =
 \Bigg[{1 \over  2 \ep^2}\Bigg\{ P^{(0)}_{qq}\otimes P^{(0)}_{qg}
 +P^{(0)}_{qg}\otimes P^{(0)}_{gg}
+2\beta_0 P^{(0)}_{qg}\Bigg\} + {1 \over  2 \epsilon} P^{(1)}_{qg}\Bigg]
\end{eqnarray}
%
In order to determine $\Delta_{ab}$, 
we set $\mu_R=\mu_F$ and expand $\Delta_{ab}$ (similarly $\hat \Delta_{ab}$) as
\begin{eqnarray}
\Delta_{ab}\left (x,\frac{m^2_{\phi}}{\mu_F^2}\right )=\sum_{i=0}^\infty a_s^i(\mu_F^2) 
\Delta_{ab}^{(i)}\left (x,\frac{m^2_{\phi}}{\mu_F^2}\right )\,.
\label{delexp}
\end{eqnarray}
Substituting the above equation (eqn. (\ref{delexp})) in
the eqn.(\ref{massfact}), we obtain to order $a_s(\mu_F^2)$
\begin{eqnarray}
\Delta_{q \overline q}^{(1)} &=& \hat \Delta_{q \overline q}^{(1)} - {2 \over \ep}
\pqqz \conv \hat \Delta_{q\overline q}^{(0)} 
\nonumber\\[2ex]
\Delta_{q g}^{(1)} &=& \hat \Delta_{q g}^{(1)} - {1 \over \ep}
\pqgz \conv  \hat \Delta_{q \overline q}^{(0)}
\label{factnlo}
\end{eqnarray}
and to order $a_s^2(\mu_R^2)$
\begin{eqnarray}
\Delta_{q\overline q}^{(2)}&=& \hat \Delta_{q\overline q}^{(2)} -2 \Gamma_{q \overline q}^{(1)}  
\conv \hat \Delta_{q \overline q}^{(0)}
-{2 \over \ep} \pqqz\conv\Delta_{q\overline q}^{(1)}
-{2 \over \ep} \pgqz\conv\Delta_{qg}^{(1)}
\nonumber\\[2ex]
&&-{1 \over \ep^2} \pqqz \conv \pqqz \conv \hat \Delta_{q\overline q}^{(0)}
\nonumber\\[2ex]
\Delta_{qg}^{(2)}&=& \hat \Delta_{qg}^{(2)} 
-\Gamma_{\overline qg}^{(1)}  
\conv \hat \Delta_{q \overline q}^{(0)}
-{1 \over \ep} \pqqz \conv \Delta_{qg}^{(1)}
-{1 \over \ep} \pqgz \conv \Delta_{q\overline q}^{(1)}
-{1 \over \ep} \pggz \conv \Delta_{q g}^{(1)}
\nonumber\\[2ex]
&&-{1 \over \ep^2} \pqqz \conv \pqgz \conv \hat \Delta_{q \overline q}^{(0)}
\nonumber\\[2ex]
\Delta_{gg}^{(2)}&=& \hat \Delta_{gg}^{(2)} 
-{4 \over \ep} \pqgz\conv\Delta_{qg}^{(1)}
-{2 \over \ep^2} \pqgz \conv \pqgz \conv\hat \Delta_{q\overline q}^{(0)}
\label{factnnlo}
\end{eqnarray}
We have computed $\hat \Delta_{ab}$ for all those
reactions that contribute to scalar production to order $a_s^2$ 
and substituted them
into eqns.(\ref{factnlo},\ref{factnnlo}) to obtain finite
partonic coefficient functions $\Delta_{ab}$.  The 
results are lengthy and hence they are presented in the Appendix
after setting $\mu_R=\mu_F$.  
It is straightforward
to obtain the $\mu_R$ dependence of the partonic coefficient functions
presented in the Appendix using the RG equations given in eqns.(\ref{RGE}). 
The mass factorisation implies DGLAP evolution equation:   
\begin{eqnarray}
\mu_F^2 {d \over d\mu_F^2}f_a(x,\mu_F^2)={1 \over 2} \sum_{b} P_{ab}
                         \left(x,\mu_F^2\right)
                        \otimes f_b (x,\mu_F^2) \,,
\label{DGLAB}
\end{eqnarray}
where $P_{ab}$ are Altarelli-Parisi splitting functions\cite{AP}.
The DGLAP evolution equation determines the scale evolution of the
parton densities appearing in eqn.(\ref{sigphi}).
Note that the RG equations of strong, Yukawa couplings (eqn.(\ref{REASLAM})) 
along with the above DGLAP evolution equations (eqn.(\ref{DGLAB})) control
both factorisation and renormalisation scale dependences of $\Delta_{ab}$
at every order in perturbation theory.

We have made several checks on our NNLO results, both analytically as well
as numerically.   First and the foremost check is the observation of cancellation of
all the poles in $\epsilon$ that result from UV, soft and collinear divergences
after all the appropriate renormalisation constants and factorisation kernels are
systematically taken into account.  The second check involves the comparison of our results
against those computed for Higgs production through bottom quark annihilation \cite{Harlander}.
Note that we have presented our results in such a way that they can be used for 
both neutral scalar (sneutrino) and charged scalar (sleptons) productions.  
So, the comparison against \cite{Harlander} is possible only after combining various 
pieces of the coefficient functions and then multiplying $x$ and setting 
$C_A=3,C_F=4/3,T_f=1/2$ in our expressions.  We found complete agreement with 
\cite{Harlander} which serves as an important check on our computation.  Finally, 
we have reproduced all those plots given in \cite{Harlander} using our code in 
order to check the correctness of numerical code.  In the next section we will discuss 
in detail the numerical impact of our results for both sneutrino and charged slepton 
productions at hadron colliders.    

Recently, there have been several breakthroughs in understanding
the structure of perturbative series to all orders in perturbation
theory, thanks to explicit results on form factors and Altarelli-Parisi
splitting functions 
(see \cite{Moch:2004pa,Vogt:2004mw,Moch:2005id,Moch:2005tm,Moch:2005ba, Blumlein:2004xt})
to three loop level in QCD.  
The computation of such quantities reveals 
the long distance physics resulting from the
soft gluon emissions in the scattering processes.
The soft gluon contributions to hadronic cross sections 
often dominate over the rest in the region where the partonic scaling 
variable $x$ approaches unity.  Due to
the peculiar behavior of the partonic fluxes in this region, often
these effects need to be resummed to all orders in perturbation theory.     
Resummation of soft gluons for hadronic
reactions can now be achieved upto next to next to next to leading logarithm ($N^3LL$) 
level using the available three loop results  
(see \cite{Moch:2005ky, Laenen:2005uz,Idilbi:2005ni,Ravindran:2005vv}).    
More on the structure of perturbative results both in fixed order as well as 
in the resummed quantities can be found in 
\cite{Blumlein:2000wh,Blumlein:2005im,Dokshitzer:2005bf,Ravindran:2005vv}.
In \cite{Ravindran:2005vv}, soft gluon enhanced partonic contributions were obtained 
for Drell-Yan, Higgs productions at hadronic colliders. 
This was achieved 
by using the collinear factorisation property of scattering cross sections, 
the Sudakov resummation of soft gluon effects and  
applying various renormalisation group techniques. 
In \cite{Ravindran:2006cg}, order $\alpha_s^3$ soft gluon contributions 
to Higgs production through bottom quark annihilation process 
at hadron colliders were presented for the first time
using the soft gluon enhanced cross sections thus obtained.
Since the coupling of SM Higgs boson to bottom quarks and 
that of sneutrino/sleptons are both of Yukawa type, we can
use $\Delta_{b \overline b}^{(3)}$ of \cite{Ravindran:2006cg} to 
study the soft gluon effects on the 
snuetrino and charged slepton production cross sections at hadron colliders.
We will present the numerical impact of these effects
towards the end of next section.   Such a study presents a quantitative estimate
on the missing higher order contributions to the processes under study.

\section{Results and Discussion}
        \label{sec:results}
Having obtained the compact analytic results for the partonic coefficient functions
$\Delta_{ab}$ for various subprocesses to NNLO in perturbative
QCD, we now endeavor to study their impacts on the 
resonant production of sneutrino and charged slepton at the LHC ($\sqrt{S}=14$ TeV)
and for the Run II of Tevatron ($\sqrt{S}=1.96$ TeV). 
As discussed in the Introduction, we will limit ourselves only to contributions from  
the first generation of quarks .   
Since at hadron colliders, the resonant production is through 
the interaction term $\lambda'_{ijk}L_iQ_jD_k^c$ in the Lagrangian (see eq.(\ref{eq:superpot})), 
for $j,k = 2,3$, the production rate will be suppressed due to the 
low flux of the sea quarks. 
To obtain the production cross section to a particular order,
one has to convolute the partonic coefficient functions $\Delta_{ab}$  with the corresponding 
parton densities $f_a$, both to the same order.   Further 
the coupling constants $a_s(\mu_R)$ and $\lambda'(\mu_R)$ 
should also be evaluated using the corresponding RGEs (eqn.(\ref{RGE})) computed to the same order.  

Following the Ref\cite{vanRitbergen:1997va},  the solution to RGE (second equation in eq.(\ref{RGE})) 
for $\lambda'(\mu_R^2)$ is given by ,
\begin{eqnarray}
\lambda'(\mu_R^2) = \lambda'(\mu_{0}^2){ C(a_s(\mu_R^2)) \over C(a_s(\mu_{0}^2))}
\end{eqnarray}
with
\begin{eqnarray}
C(a_s) &=& a_s^{A_0} \sum_{i=0}^\infty a_s^i~C_i\,.
\end{eqnarray}
The $C_i$ are given by 
\begin{eqnarray}
C_0&=&1,\quad \quad C_1=A_1,\quad \quad
\nonumber\\[2ex]
C_2&=&{1 \over 2} (A_1^2+A_2),
\quad \quad C_3={1 \over 6} (A_1^3+3 A_1 A_2+2 A_3),
\end{eqnarray}
with
\begin{eqnarray}
      A_0&=&c_0, \quad \quad
      A_1=c_1-b_1 c_0, \quad \quad
      A_2=c_2-b_1 c_1+c_0 (b_1^2-b_2),
\nonumber\\[2ex]
      A_3&=&c_3-b_1 c_2+c_1 (b_1^2-b_2)+c_0 (b_1 b_2-b_1 (b_1^2-b_2)-b_3),
\end{eqnarray}
and 
\begin{eqnarray}
c_i={\gamma_i \over \beta_0},\quad \quad  b_i={\gamma_i \over \beta_0},
\end{eqnarray}
and $\mu_{0}$ is some reference scale at which both $a_s$ as well as $\lambda'$ are
known.
We have numerically evaluated $a_s(\mu_R^2)$ and $\lambda'(\mu_R^2)$ to relevant order 
namely LO, NLO and NNLO 
by truncating the terms in the RHS of eqn.(\ref{REASLAM}). We have used 
$\lambda'(M_Z^2)= 0.01$ irrespective of flavor and mass of the sneutrino/charged slepton.
We have used the latest MSTW parton densities \cite{MSTW2008} in our numerical code and the
corresponding values of $\alpha_s(M_Z)$ for LO, NLO and NNLO provided with
the sets.

\subsection{Sneutrino Production}

The total sneutrino production cross section as function of its mass
is plotted in fig.\ \ref{fig:l0} for LHC (left panel) and Run II of Tevatron (right panel) energies.  
We have set the renormalisation scale to be the mass of the
sneutrino, $\mu_{R}=m_{\tilde{\nu}}$.
The pair of lines corresponds to the two extreme choices of factorisation scale: 
$\mu_{F}=m_{\tilde{\nu}}$ (upper) and $\mu_{F}= 0.1 m_{\tilde{\nu}}$ (lower). 
The plots clearly demonstrate that the NNLO contributions reduce 
the factorisation scale dependence improving the theoretical predictions
for sneutrino production cross section.  To quantify the percentage variation
with respect to the factorisation scale we define 
\begin{eqnarray}
\delta \sigma^\phi_{I}(\mu_F) &=& {\sigma^\phi_{I}\left(\mu_F=\mu_R=m_{\tilde \nu}\right) 
- \sigma^\phi_{I}\left(\mu_F=0.1 m_{\tilde \nu},\mu_R=m_{\tilde \nu}\right) \over 
\sigma^\phi_{I}\left(\mu_F=\mu_R=m_{\tilde \nu}\right)
}
\end{eqnarray}
where $I=LO,NLO,NNLO$.  
\begin{table}[htb]
\begin{center}
\bigskip
\begin{tabular}{||c|c|c|c|c||}
\hline
 & Mass Range $(m_{\tilde \nu})$& LO in \%& NLO in \%& NNLO in \%\\[1ex]
\hline
\hline
Tevatron & 100 GeV to 1 TeV & $- 4.6 $ to $140 $ & $-2.2 $ to $-39$ & 
$-1.1 $ to $-10.6 $ \\[1ex]
\hline
LHC & 100 GeV to 450 GeV & $ 40 $ to $2.3  $ & $ 23 $ to $ 1.8 $ & 
$~~9.3 $ to $1.1 $ \\[1ex]
    & 510 GeV to 1 TeV & $- 1.2 $ to $-17$   & $- 0.22 $ to $- 5.2 $ & 
$-0.24 $ to $-1.2 $ \\[1ex]
\hline
\end{tabular}
\caption[] {$\delta \sigma^{(i)}(\mu_F)$ in $\%$ in the mass range between $100~$GeV and
$1~$TeV at Tevatron and LHC energies.
}
\label{tab:muf_var}
\end{center}
\end{table}
The cross section falls off with the sneutrino
mass due to the availability of phase space with respect to the mass,
the choice of $\mu_R=m_{\tilde{\nu}}$ and the parton densities.  The latter effect, 
understandably, is more pronounced at the Tevatron than at the LHC. 
 
In order to estimate the magnitude of the QCD corrections at NLO and NNLO, we 
define the K-factors as follows: 
$$K^{(1)} = \sigma^\phi_{\rm NLO} / \sigma^\phi_{\rm LO} \hspace{0.5cm} 
K^{(2)} = \sigma^\phi_{\rm NNLO} / \sigma^\phi_{\rm LO}.$$ 
In fig.\ref{fig:kl0}, we have plotted both $K^{(i)}$ ($i=1,2$) as a function
of sneuttrino mass.  We have chosen $\mu_F=
\mu_R=m_{\tilde \nu}$ for this study. 
At the LHC, The $K^{(1)}$ varies between $1.23$ to $1.46$ and $K^{(2)}$ between
$1.27$ to $1.52$ in the mass range $100~GeV \le m_{\title \nu} \le 1~TeV$.  
At the Tevatron, we find that $K^{(1)}$ varies between $1.55$ to $1.53$ and $K^{(2)}$ between
$1.65$ to $1.85$ for the same mass range.
Note that numbers for $K^{(1)}$ differ slightly from those given in our earlier work \cite{swapan}
due to the running of $\lambda'$ in the present analysis.  
We also observe that $K$ factor is much bigger at the Tevatron compared to that of at the LHC. 
The reason behind this is attributed to the different behavior of parton densities at the Tevatron and the LHC.
Note that parton densities rise steeply as $x \rightarrow 0$ and fall off very fast as $x \rightarrow 1$, 
which means the dominant contribution to the production results from the phase space region where 
$x \sim \tau (= m^2_{\tilde \nu}/S)$ becomes small.
$\tau$ at Tevatron ($0.05\lsim\tau \lsim 0.5$) is larger compared to
that at LHC ( $0.007\lsim\tau\lsim 0.07$) (see also fig.\ref{fig:kl0}). 
Because of this, at Tevatron the valence quark initiated processes dominate while 
gluon and sea quark initiated processes dominate at the LHC. 
As the mass of the sneutrino increases, that is $x$ approaches to unity,
the $K$-factor at Tevatron naturally falls off. 
At LHC, in the higher mass region ($\sim 1$ TeV),
valence quark densities start to dominate and hence it stays 
almost flat compared to Tevatron. 
We find that the most dominant 
sub processes are the $d\bar d$ and $dg$ initiated processes.  The $dg$
sub process which begins at NLO gives a negative contribution both 
at NLO and NNLO. 

We now turn to study the impact of   
the factorisation scale $(\mu_F)$ and 
the renormalisation scale ($\mu_R$) on the production cross section. 
The factorisation scale dependence for both LHC (left panel) 
and Tevatron (right panel) are shown in upper panels of fig.\ \ref{fig:ml0}, 
for $m_{\tilde \nu}=300~GeV$ (LHC), $m_{\tilde \nu}=120~GeV$ (Tevatron).  We have 
chosen $\mu_R=m_{\tilde \nu}$ for both the LHC and the Tavatron.   
The factorisation scale 
is varied between $\mu_F=0.1 ~m_{\tilde \nu}$ and $\mu_F=10 ~m_{\tilde \nu}$.  
We find that the factorisation scale dependence decreases 
in going from LO to NLO to NNLO as expected.   

The dependence of the renormalisation scale dependence on the total
cross sections for the resonant production of sneutrino at the LHC and the Tevatron
is shown in the lower panels of fig.\ \ref{fig:ml0}.
Note that the LO is already $\mu_R$ dependent  
due to the coupling $\lambda' (\mu_R)$. 
We have performed this analysis for sneutrino mass
$m_{\tilde{\nu}}\,
= 300~GeV$ (LHC), $m_{\tilde \nu}=120~GeV$ (Tevatron).  
We have set the factorisation scale $\mu_F=m_{\tilde \nu}/4$ 
(see for example, Ref.\cite{choos_fact_scal} for more details) and
the renormalisation scale is varied in the  
range $0.1 \le \mu_R/m_{\tilde \nu} \le 10$. 
We find significant 
reduction in the $\mu_R$ scale dependence when higher
order QCD corrections are included. 
It is clear from both the panels of fig.\ \ref{fig:ml0}  that
our present NNLO result makes the predictions almost independent
of both factorisation and renormalisation scales.
\subsection{Charged slepton production}
We now study the numerical impact of our NNLO results on the charged slepton production
for both Tevatron and LHC energies. 
In fig. \ref{fig:lp}, we have plotted the total cross section as function of 
charged slepton mass.  The upper (lower) set of lines corresponds to the
factorisation scale $\mu_F = m_{\tilde{\ell}^+}(0.1 m_{\tilde{\ell}^+})$. 
The improvement due to NLO and NNLO pieces is evident.
As we expect 
the $K$-factors (see figs.\ref{fig:klp}) are quite similar to the case of neutral scalar production.
In upper panels of fig. (\ref{fig:mlp}), we have shown 
the production cross section as a function of factorisation scale $\mu_F$
for the   
slepton mass $m_{\tilde{\ell^+}} = 300 GeV$ (LHC),
$m_{\tilde{\ell^+}} = 120 GeV$ (Tevatron) 
and fixed the renormalisation scale 
at $\mu_R=m_{\tilde{\ell^+}}$.
In the lower panels of fig. (\ref{fig:mlp}), we have shown 
$\mu_R$ variation for the LHC and the Tevatron. 
We have done this for 
charged slepton mass $m_{\tilde{\ell^+}} = 300$ GeV (LHC),
$m_{\tilde{\ell^+}} = 120$ GeV (Tevatron)
and fixed the factorisation scale 
$\mu_F = {1 \over 4}\, m_{\tilde{\ell^+}}$ (see Ref.\cite{choos_fact_scal}).
We again find that the scale dependence gets reduced significantly as we include
higher order terms in the perturbative expansion.
Our numerical code can also produce results for  
$\tilde \ell^-$ production both at the LHC and the Tevatron. 
At Tevatron, the production rates for $\tilde \ell^-$ is found to be same as that of
$\tilde \ell^+$ because the contributing fluxes and the partonic coefficient functions
are identical.   At the LHC, the is not case because of different fluxes that contribute
for $\tilde \ell^+$ and $\tilde \ell^-$.

\subsection{Soft gluon contributions at order $\alpha_s^3$}
The order $\alpha_s^3$ partonic coefficient function $\Delta_{b\overline b}^{(3)}$
resulting from the soft gluons is
known for the Higgs boson production through $b\overline b$ annihilation 
(see \cite{Ravindran:2006cg}).  
The same coefficient function can be used here to study the impact of 
soft gluons for the sneutrino (also
for the charged slepton) production at $\alpha_s^3$ level due to identical structure of the
interaction terms responsible for their production mechanisms. 
For our numerical study, we have used the available 
three loop $\beta_3$ and $\gamma_3$ to evolve
$\alpha_s$ and $\lambda'$ respectively.  For parton density sets, we can only use 
the available NNLO evolved MSTW 2008 sets.  In fig. (\ref{fig:n3l0}), we have
plotted the $N^3LO$ corrected sneutrino production cross sections 
against its mass (upper panels)
and the renormalisation scale (lower panels) on for both LHC as well as 
Tevatron energies.  While the order $\alpha_s^3$ soft gluon effects  
to the production of sneutrinos is indistinguishable with respect to exact $NNLO$ 
contributions, it reduces the uncertainty resulting from the choice of 
renormalisation scale significantly.  

\section{Conclusions}
        \label{sec:conclusions}
The ongoing program in search of signals of BSM scenarios
can be successful only if the theory predictions are precise
and reliable.  In this paper we have attempted to
make predictions for the production of sneutrinos and sleptons
at hadron colliders such as Tevatron and the LHC.  The resonant
production of sneutrinos and sleptons are possible 
at these colliders thanks to $R$-parity violating interactions present 
in the supersymmetric theory which is one of the most studied
BSM in the literature.
Often predictions based on subprocess contributions computed
at leading order in perturbation theory suffer from uncertainties
resulting from the arbitrariness in the choice of 
renormalisation and factorisation scales.  These scales are artifacts
of the perturbation theory and hence are unphysical.
The sensitivity to these scales signals the missing higher order
contributions that need to be included in order to make the predictions reliable. 
In other words, contributions from NLO and NNLO sub processes are expected
to reduce these theoretical uncertainties.  In addition,
the potential SM background processes to resonant production of sneutrinos and sleptons and 
their subsequent decays are well under control as they are known to NNLO level in QCD.
Hence, we have computed all the subprocess contributions to
the production cross sections for sneutrino and sleptons upto order $\alpha_s^2$ (i.e., NNLO) 
in perturbative QCD.  We have used dimensional regularisation to regulate
UV, soft and collinear divergences.  The renormalisation and the factorisation  
are done in ${\overline {MS}}$ scheme.  We have used the latest parton density sets provided by 
MSTW for all our analysis.  We have demonstrated how the inclusion of
NNLO contributions can reduce the scale uncertainties over wide range of
sneutrino and slepton masses for both Tevatron and LHC energies.   
We also find significant increase in the cross section due to opening up of
several new partonic channels beyond the leading order.  The increase
in the cross section due to the inclusion of NNLO contributions compared to that of NLO 
is found to be $6.6\%$ to $21\%$ for Tevatron and $3.4\%$ to $4\%$ for LHC 
and the total NNLO K factor varies from $1.65$ to $1.85$ for Tevatron and $1.27$ to $1.52$ for
LHC in the mass range of $100~$ GeV to $1~$ TeV. 
In order to estimate
the impact of QCD corrections beyond NNLO, we have performed an analysis
taking into account the dominant soft gluon contributions at $N^3LO$ and found
that they are stable under perturbation. 
The calculations presented in this paper are not particular to
supersymmetric theories, but can be applied to any
color-singlet scalar (pseudoscalar) coupling to a quark anti-quark pair.

\section*{Acknowledgements}
The work of V.R. has been partially supported by funds made
available to the Regional Centre for Accelerator based Particle Physics
(RECAPP) by the Department of Atomic Energy, Govt. of India.  We would
like to thank the cluster computing facility at Harish-Chandra Research
Institute where part of computational work for this study was carried
out.  S.M would like to thank RECAPP center for his visit, where part of
the work was done.

\newpage
\section{Appendix}
The leading order contribution (fig.(\ref{fig1})) from the subprocess $q_i +\overline q_j 
\rightarrow \phi$ gives:
\begin{eqnarray}
   \hat \Delta^{(0)} _{q\overline q} =
   \Delta^{(0)} _{q\overline q} =
       \deltaOmx 
\end{eqnarray}
To order $a_s$, we need to include 
$q_i +\overline q_j \rightarrow g + \phi$,
$q_i(\overline q_i) +g \rightarrow q_j(\overline q_j) + \phi$ 
and one loop corrections to $q_i +\overline q_j \rightarrow \phi$. The 
quark anti-quark initiated processes (fig.(\ref{fig2},\ref{fig3})) give 
\begin{eqnarray}
   \Delta^{(1)} _{q\overline q} &=&
        C_F  \Bigg[ 
         2 - 2 x 
       + \logx   \Bigg\{ 2 - {4 \over (1-x)} + 2 x \Bigg\}
       + \logONmx   \{  - 4 - 4 x \}
\nonumber\\[2ex]
&&       + 8\left({\logONmx \over (1-x)} \right)_+
       + \logqmuf   \Bigg\{  - 2 + {4\over (1-x)_+} - 2 x \Bigg\}
       + \deltaOmx   \Big\{  - 2 + 4 \zeta_2 \Big\}
        \Bigg]
\nonumber\\[2ex]
&&       + \ep~ C_F \Bigg[\Bigg(  - {3 \over (1-x)_+} \zeta_2 
+ {3 \over 2} \zeta_2 + {3\over 2} x \zeta_2 \Bigg)
       +  \logx   \Big\{  - 1 + x \Big\}
\nonumber\\[2ex]
&&       +  \logTWx   \Bigg\{  - {1 \over 2} + {1 \over (1-x)} - {1\over 2} x \Bigg\}
       +  \logONmx   \Bigg\{ 2 - 2 x 
       +  \logx   
\nonumber\\[2ex]
&&      \times  \Bigg( 2 - {4 \over (1-x)} + 2 x \Bigg)\Bigg\}
       +  \logTWOmx   \{  - 2 - 2 x \}
+ 4 \left({ \logTWOmx \over (1-x)}\right)_+ 
\nonumber\\[2ex]
&&       +  \logqmuf   \Bigg\{ 1 - x 
       +  \logx   \Bigg(1 - {2 \over (1-x)} + x \Bigg)
       +  \logONmx   
       (  - 2 - 2 x)
\nonumber\\[2ex]
&&         + 4 \left({ \logONmx \over (1-x)}\right)_+ 
\Bigg\}
       +  \logTWqmuf   \Bigg\{  - {1 \over 2} + {1 \over (1-x)_+} - {1 \over 2} x \Bigg\}
\nonumber\\[2ex]
&&       +  \deltaOmx   \Bigg\{ 2 
       +  \logqmuf   \Big(  - 1 + 2 \zeta_2 \Big) \Bigg\}
         \Bigg]
\end{eqnarray}
and quark (anti-quark) gluon initiated processes (fig.(\ref{fig4})) give
\begin{eqnarray}
   \Delta^{(1)} _{qg} &=& \Delta^{(1)} _{\overline qg} 
\nonumber\\[2ex]
&=& T_f \Bigg[
           - {3 \over 2} + 5 x - {7 \over 2} x^2 
       + \logx   \Big\{  - 1 + 2 x - 2 x^2 \Big\}
\nonumber\\[2ex]
&&       + \logONmx   \Big\{ 2 - 4 x + 4 x^2 \Big\}
       + \logqmuf   \Big\{ 1 - 2 x + 2 x^2 \Big\}
         \Bigg]
\nonumber\\[2ex]
&&       +\ep~ T_f \Bigg[ {3 \over 2} - {3 \over 4} \zeta_2 - 5 x + {3 \over 2} x \zeta_2
          + {7 \over 2} x^2 - {3 \over 2} x^2 \zeta_2 
       + \logx   \Bigg\{ {3 \over 4} - {5 \over 2} x + {7 \over 4} x^2 \Bigg\}
\nonumber\\[2ex]
&&       + \logTWx   \Bigg\{ {1 \over 4} - {1 \over 2} x + {1 \over 2} x^2 \Bigg\}
       + \logONmx   \Bigg\{  - {3 \over 2} + 5 x - {7 \over 2} x^2 
\nonumber\\[2ex]
&&       + \logx   \Big(  - 1 + 2 x - 2 x^2 \Big) \Bigg\}
       + \logTWOmx   \Big\{ 1 - 2 x + 2 x^2 \Big\}
\nonumber\\[2ex]
&&       + \logqmuf   \Bigg\{  - {3 \over 4} + {5 \over 2} x - {7 \over 4} x^2 
       + \logx   \Bigg(  - {1 \over 2} + x - x^2 \Bigg)
\nonumber\\[2ex]
&&       + \logONmx   \Big( 1 - 2 x + 2 x^2 \Big)\Bigg\}
       + \logTWqmuf   \Bigg\{ {1 \over 4} - {1 \over 2} x + {1 \over 2} x^2 \Bigg\}
       \Bigg]
\end{eqnarray}
In the above results we have presented results to order $\ep$ that will contribute to
NNLO restults.   

To order $a_s^2$, several partonic subprocesses contribute.  We present 
the contributions coming from each subprocess below.  
The quark anti-quark initiated processes with no gluon (fig.(\ref{fig5})), one gluon 
(fig.(\ref{fig6})) and two gluons (fig.(\ref{fig7})) in the final
state along with $\phi$ constitute a sub class giving order $a_s^2$ contribution.
This class consists of processes with two loop corrections to leading order 
$q_i +\overline q_j \rightarrow \phi$ (no gluon)
along with one loop corrections to $q_i +\overline q_j \rightarrow \phi +g$ (one gluon).  In addition 
we have $q_i +\overline q_j \rightarrow g + g + \phi$ (two gluons) in this class.
The unrenormalised form factor ${\cal F}_\phi$ upto two loop order (fig.(\ref{fig5})) is found to be
\cite{Ravindran:2006cg}
\begin{eqnarray}
{\cal F}_\phi\left(m_\phi^2,\mu^2\right)&=&1+
\hat a_s \left(-{m_\phi^2 \over \mu^2}\right)^{\ep \over 2} { S}_{\ep}
  C_F \Bigg\{-{8\over \ep^2}-2+\zeta_2
               +\ep \Bigg(2-{7 \over 3} \zeta_3\Bigg)
               +\ep^2 \Bigg(-2+{1\over 4} \zeta_2
\nonumber\\[2ex]
&&+{47\over 80} \zeta_2^2\Bigg)\Bigg\}
+\hat a_s^2 \left(-{m_\phi^2 \over \mu^2}\right)^{\ep} {S}^2_{\ep}
       \Bigg[ C_A C_F  \Bigg\{  - {1655 \over 81} + {44 \over 3 \ep^3} + {4 \over \ep^2}
         \zeta_2 - {134 \over 9 \ep^2}
\nonumber\\[2ex]
&&+ {11 \over 3 \ep}\zeta_2 - {26\over
         \ep} \zeta_3 + {440 \over 27 \ep} - {103 \over 18} \zeta_2 + {44 \over 5}
         \zeta_2^2 + {305 \over 9} \zeta_3 \Bigg\}
       + n_f C_F   \Bigg\{ {416 \over 81} - {8\over 3 \ep^3 }
\nonumber\\[2ex]
&&+ {20 \over 9 \ep^2} - {2\over 3
         \ep}\zeta_2 - {92 \over 27 \ep} + {5 \over 9}\zeta_2 - {26 \over 9}
         \zeta_3 \Bigg\}
       + C_F^2  \Bigg\{ 22 + {32 \over \ep^4}- {8 \over \ep^2}\zeta_2 + {16 \over
         \ep^2}
\nonumber\\[2ex]
&&- {12 \over \ep}\zeta_2 + {128 \over 3 \ep}\zeta_3 - {16 \over
         \ep} + 12\zeta_2 - 13\zeta_2^2 - 30\zeta_3 \Bigg\}\Bigg]
\end{eqnarray}
The contributions from quark antiquark annihilation processes can be split into two parts: 
contributions coming from threshold region
called soft plus virtual (S+V) contribution: 
\begin{eqnarray}
   \Delta^{(2),S+V} _{q\overline q} &=& 
        n_f C_F       
       \Bigg[
         {32 \over 3} \Dtwo(x)  - {160 \over 9}  \Done(x)  + {224 \over 27} 
         \Dzero(x)  - {32 \over 3}  \Dzero(x)   \zeta_2  
\nonumber\\[2ex]
&&       + \logqmuf    
       \Bigg\{ {32 \over 3}  \Done(x)  
      - {80 \over 9}  \Dzero (x)  
                     \Bigg\}
       + \logTWqmuf    \Bigg\{ {8 \over 3}  \Dzero(x)  \Bigg\}
\nonumber\\[2ex]
&&       + \deltaOmx    \Bigg\{ {8 \over 9} + 8  \zeta_3  - {40 \zeta_2\over 9} \Bigg\}
       \Bigg]
       + C_F ^2   
       \Bigg[
          128 \Dthree(x)  - 64  \Done(x)  - 128  \Done(x)  
         \zeta_2  
\nonumber\\[2ex]
&&       + 256  \Dzero(x)   \zeta_3  
       + \logqmuf    \Bigg\{ 192  \Dtwo(x)  - 32  \Dzero(x)  - 64 
         \Dzero(x)   \zeta_2  \Bigg\}
\nonumber\\[2ex]
&&       + \logTWqmuf    \Big\{ 64  \Done(x)  \Big\}
       + \deltaOmx \Bigg(    16 - 60  \zeta_3  + {8 \over 5}  \zeta_2 ^2 
       +    \logqmuf    \Big\{ 176  \zeta_3  
\nonumber\\[2ex]
&&- 24  \zeta_2  \Big\}
       +   \logTWqmuf    \Big\{  - 32  \zeta_2  \Big\} \Bigg)
       \Bigg]
       + C_A   C_F    
       \Bigg[
           - {176 \over 3} \Dtwo(x)  + {1072 \over 9}  \Done(x)  
\nonumber\\[2ex]
&&       - 32 \Done(x)   \zeta_2  
- {1616 \over 27}  \Dzero(x)  + 56  \Dzero(x)   \zeta_3 
          + {176 \over 3} \Dzero(x)   \zeta_2  
       + \logqmuf    
\nonumber\\[2ex]
&&     \times \Bigg\{  - {176 \over 3}  \Done(x)  
       + {536 \over 9}  \Dzero(x)  - 
         16 \Dzero(x)   \zeta_2  \Bigg\}
       + \logTWqmuf    \Bigg\{  - {44 \over 3}  \Dzero(x)  \Bigg\}
\nonumber\\[2ex]
&&       + \deltaOmx    \Bigg( {166 \over 9} - 8  \zeta_3  + {232 \over 9}  \zeta_2  - 
         {12 \over 5} \zeta_2 ^2 
       +  \logqmuf    \Big\{  - 12 - 24  \zeta_3  \Big\} \Bigg)
        \Bigg]
\end{eqnarray}
where the "plus" distributions ${\cal D}_i(x)$ are given by 
\begin{eqnarray}
{\cal D}_i(x)&=&\left(\log^i(1-x) \over 1-x\right)_+
\nonumber
\end{eqnarray}
and the hard contribution whose $C_A C_F$ part is given by 
\begin{eqnarray}
   \Delta^{(2),C_A} _{q\overline q} &=&
        C_A   C_F    
       \Bigg[
           {1186 \over 27} - 28 \zeta_3  - {100 \over 3}  \zeta_2  + {430 \over 27} x
          - 28 x \zeta_3  - {76 \over 3} x  \zeta_2  
\nonumber\\[2ex]
&&       + \sONTWOmx    \Bigg\{ 4 - {8 \over  (1-x) } + 4 x \Bigg\}
        + \liTHOmx    \Bigg\{ 12 - {24 \over  (1-x) } + 12 x \Bigg\}
\nonumber\\[2ex]
&&      + \liTWONmx    \Bigg\{  - 16 + {8 \over 3  (1-x) } - 4 x \Bigg\}
       + \logx    \Bigg\{ 40 - {232 \over 3  (1-x) } 
\nonumber\\[2ex]
&&    + {16 \over  (1-x) }  \zeta_2 
          - 8 \zeta_2  + {154 \over 3} x - 8 x  \zeta_2  \Bigg\}
       + \logx   \liTWONmx    \Bigg\{  - 8 
\nonumber\\[2ex]
&&    + {16 \over  (1-x) } - 8 x \Bigg\}
       + \logTWx    \Bigg\{ {59 \over 3} - {29 \over  (1-x) } + {65 \over 3} x \Bigg\}
       + \logONmx    
\nonumber\\[3ex]
&&    \times \Bigg\{  - {416 \over 9} + 16  \zeta_2  - {692 \over 9} x + 16 x 
         \zeta_2  \Bigg\}
       + \logONmx   \liTWONmx  
\nonumber\\[2ex]
&&     \times  \Bigg\{ 8 - {16 \over  (1-x) } + 8 x \Bigg\}
       + \logONmx   \logx    \Bigg\{  - {176 \over 3} + {280 \over 3  (1-x) } - {176 \over 3} x
          \Bigg\}
\nonumber\\[2ex]
&&       + \logTWOmx    \Bigg\{ {88 \over 3} + {88 \over 3} x \Bigg\}
       + \logqmuf \Bigg(   \Bigg\{  - {208 \over 9} + 8  \zeta_2  - {364 \over 9} x + 8 x 
         \zeta_2  \Bigg\}
\nonumber\\[2ex]
&&       +  \liTWONmx    \Bigg\{ 8 - {16 \over  (1-x) } + 8 x \Bigg\}
       +   \logx    \Bigg\{  - {88 \over 3} + {140 \over 3  (1-x) } - {88 \over 3} x
          \Bigg\}
\nonumber\\[2ex]
&&       +  \logONmx    \Bigg\{ {88 \over 3} + {88 \over 3} x \Bigg\}\Bigg)
       + \logTWqmuf    \Bigg\{ {22 \over 3} + {22 \over 3} x \Bigg\}
       \Bigg]
\end{eqnarray}
and the $C_F^2$ part is given by 
\begin{eqnarray}
   \Delta^{(2),C_F} _{q\overline q} &=&
        C_F ^2   
       \Bigg[
            - 72 - 128 \zeta_3  + 96  \zeta_2  + 64 x - 128 x 
         \zeta_3  - 96 x  \zeta_2  
       + \sONTWOmx 
\nonumber\\[2ex]
&&     \times   \Bigg( 80 - {64 \over  (1-x) } + 80 x \Bigg)
       + \liTHOmx    \Bigg(  - 64 - {16 \over  (1-x) } - 64 x \Bigg)
\nonumber\\[2ex]
&&       + \liTWONmx    (  - 24 - 16 x )
       + \logx \Bigg\{     - 56 + {32 \over  (1-x) } + {128 \over  (1-x) }  \zeta_2  
\nonumber\\[2ex]
&&       - 96
          \zeta_2  - 96 x  \zeta_2  
       + \liTWONmx    \Bigg( 32 - {48 \over  (1-x) } + 32 x \Bigg) \Bigg\}
\nonumber\\[2ex]
&&       + \logTWx    \{  - 8 + 16 x \}
       + \logTHx    \Bigg\{ {58 \over 3} - {24 \over  (1-x) } + {58 \over 3} x \Bigg\}
       + \logONmx 
\nonumber\\[2ex]
&&       \times \Bigg\{    80 + 64  \zeta_2  - 4 x + 64 x  \zeta_2  
       + \liTWONmx    \Bigg( 40 + {48 \over  (1-x) } + 40 x \Bigg)
\nonumber\\[2ex]
&&       + \logx    \{ 24 - 56 x \}
       + \logTWx    \Bigg(  - 96 + {144 \over  (1-x) } - 96 x \Bigg)\Bigg\}
       + \logTWOmx 
\nonumber\\[2ex]
&&     \times  \Bigg\{ (  - 32 + 32 x )
       + \logx    \Bigg( 156 - {248 \over  (1-x) } + 156 x \Bigg) \Bigg\}
\nonumber\\[2ex]
&&       + \logTHOmx    \{  - 64 - 64 x \}
       + \logqmuf\Bigg\{     40 + 32  \zeta_2  + 32 x  \zeta_2  
       + \liTWONmx 
\nonumber\\[2ex]
&&     \times   \Bigg( 16 + {32 \over  (1-x) } + 16 x \Bigg)
       + \logx    \{ 16 - 32 x \}
       + \logTWx    \Bigg\{  - 36 + {48 \over  (1-x) } 
\nonumber\\[2ex]
&&      - 36 x \Bigg\}
       + \logONmx    \{  - 32 + 32 x \}
       + \logONmx   \logx    \Bigg\{ 144 - {224 \over  (1-x) } 
\nonumber\\[2ex]
&&     + 144 x \Bigg\}
       + \logTWOmx    (  - 96 - 96 x ) \Bigg\}
       + \logTWqmuf \Bigg\{     - 16 + 16 x 
\nonumber\\[2ex]
&&       + \logx    \Bigg\{ 24 - {32 \over  (1-x) } + 24 x \Bigg\}
       + \logONmx    (  - 32 - 32 x )\Bigg\}
       \Bigg]
\end{eqnarray}
The functions $S_{np}$ and $Li_n$ are Nielsen integral and polylogarithm
respectively:
\begin{eqnarray}
S_{np}(x)&=&{(-1)^{n+p-1} \over (n-1)!~ p! } \int_0^1 d z {\log^{n-1}(z) \log^p(1-xz) \over z},
\nonumber\\[2ex]
Li_n(x)&=& S_{n-1,1}(x),\,\quad \quad 
\zeta_s=\sum_{n=1}^\infty n^{-s}
\end{eqnarray}
and for $s=2,3$, we have  $\zeta_2 = \pi^2/6, \zeta_3=1.202056903159\cdot \cdot \cdot $.
In addition, quark anti-quark pair in the initial and final states 
along with $\phi$ (A and B in fig. (\ref{fig8})) 
and quark (anti-quark) quark (anti-quark) pair
in the final state with $\phi$ also arise at order $a_s^2$ (C and D 
in fig. (\ref{fig9})).  
The contributions coming from the later will depend on whether we have identical
quarks (anti-quarks) or non-identical quarks (anti-quarks).  We present their contributions
below along with various allowed interferences.

The $s$-channel processes (A in fig.(\ref{fig8})) with $\phi$ emitted from incoming partons give
\begin{eqnarray}
   \Delta^{(2),A\overline A} _{q\overline q} &=&
        n_f C_F   
       \Bigg[
           - {4 \over 27} + {16 \over 3} \zeta_2  - {220 \over 27} x + {16 \over 3} x
         \zeta_2
       + \liTWONmx    \Bigg(  - {8 \over 3  (1-x) } \Bigg)
\nonumber\\[2ex]
&&       + \logx    \Bigg\{  - 4 + {40 \over 3  (1-x) } - {28 \over 3} x \Bigg\}
       + \logTWx    \Bigg\{  - {14 \over 3} + {8 \over  (1-x) } - {14 \over 3} x \Bigg\}
\nonumber\\[2ex]
&&       + \logONmx    \Bigg\{ {32 \over 9} + {128 \over 9} x
       + \logx    \Bigg( {32 \over 3} - {64 \over 3  (1-x) } + {32 \over 3} x \Bigg) \Bigg\}
\nonumber\\[2ex]
&&       + \logTWOmx    \Bigg\{  - {16 \over 3} - {16 \over 3} x \Bigg\}
       + \logqmuf    \Bigg\{ {16 \over 9} + {64 \over 9} x
       + \logx
\nonumber\\[2ex]
&&      \times \Bigg( {16 \over 3} - {32 \over 3  (1-x) } + {16 \over 3} x \Bigg)
       + \logONmx    \Bigg(  - {16 \over 3} - {16 \over 3} x \Bigg) \Bigg\}
\nonumber\\[2ex]
&&       + \logTWqmuf    \Bigg\{  - {4 \over 3} - {4 \over 3} x \Bigg\}
         \Bigg]
\end{eqnarray}
The interference of $s$-channel processes (A in fig.(\ref{fig8})) with $\phi$ emitted from 
incoming partons and $t$-channel processes (C and D in fig.(\ref{fig9})) give
\begin{eqnarray}
   \Delta^{(2),A\overline C} _{q\overline q} &=&
   \Delta^{(2),A\overline D} _{q\overline q} 
\nonumber\\[2ex]
&=& \Bigg(C_F-{C_A\over 2}\Bigg)   C_F    
       \Bigg[
          94 - 86 x 
       + \sONTWOmx    \Bigg( 28 - {72 \over  (1-x) } + 28 x \Bigg)
\nonumber\\[2ex]
&&       + \liTHOmx    \Bigg(  - 16 + {32 \over  (1-x) } - 16 x \Bigg)
       + \liTWONmx    \Bigg(  - 32 - {12 \over  (1-x) } 
\nonumber\\[2ex]
&&     - 24 x \Bigg)
       + \logx    \Bigg\{ 36 + {16 \over  (1-x) } - 30 x 
       + \liTWONmx    \Bigg( 8 - {24 \over  (1-x) } 
\nonumber\\[2ex]
&&     + 8 x \Bigg)\Bigg\}
       + \logTWx    \Bigg\{  - 2 + {15 \over  (1-x) } + 2 x \Bigg\}
       + \logTHx    \Bigg\{  - {10 \over 3} + {16 \over 3  (1-x) } 
\nonumber\\[2ex]
&&     - {10 \over 3} x \Bigg\}
       + \logONmx    \Bigg\{  - 64 + 56 x 
       + \liTWONmx    \Bigg( 16 - {32 \over  (1-x) } + 16 x \Bigg)
\nonumber\\[2ex]
&&       + \logx    \Bigg(  - 16 - {24 \over  (1-x) } - 16 x \Bigg)
       + \logTWx    \Bigg( 8 - {16 \over  (1-x) } + 8 x \Bigg) \Bigg\}
\nonumber\\[2ex]
&&       + \logqmuf   \Bigg\{   - 32 + 28 x 
       + \liTWONmx    \Bigg( 8 - {16 \over  (1-x) } + 8 x \Bigg)
\nonumber\\[2ex]
&&       + \logx    \Bigg(  - 8 - {12 \over  (1-x) } - 8 x \Bigg)
       + \logTWx    \Bigg( 4 - {8 \over  (1-x) } + 4 x \Bigg) \Bigg\}
       \Bigg]
\end{eqnarray}
The $s$-channel processes (B in fig.(\ref{fig8})) with $\phi$ emitted from outgoing partons give
\begin{eqnarray}
   \Delta^{((2),B\overline B} _{q\overline q} &=&
        C_F   T_f    
       \Bigg[
            - {16 \over 3} + {64 \over 3} x - 16 x^2 - {16 \over 3} x^2 \zeta_2  
       + \liTWmx    \Bigg(  - {32 \over 3} x^2 \Bigg)
\nonumber\\[2ex]
&&       + \logx    \Bigg\{  - {8 \over 3} + {16 \over 3} x + 8 x^2 \Bigg\}
       + \logTWx    \Bigg\{ {8 \over 3} x^2 \Bigg\}
\nonumber\\[2ex]
&&       + \logONpx   \logx    \Bigg\{  - {32 \over 3} x^2 \Bigg\}
       \Bigg]
\end{eqnarray}
The interference of $s$-channel processes (B in fig.(\ref{fig8})) with $\phi$ emitted from 
outgoing partons and $t$-channel processes (C and D in fig.(\ref{fig9})) give
\begin{eqnarray}
   \Delta^{(2),B \overline C} _{q\overline q} &=&
   \Delta^{(2),B \overline D} _{q\overline q} 
\nonumber\\[2ex]
    &=&
        \Bigg(C_F-{C_A \over 2 }\Bigg)  C_F    
       \Bigg[
          - 6 - 4 \zeta_2  + 32 x + 8 x  \zeta_2  - 26 
         x^2 
       + \sONTWmx    \Big(  - 48 x^2 \Big)
\nonumber\\[2ex]
&&       + \sONTWOmx    \Big( 32 x^2 \Big)
       + \liTHmx    \Big(  - 8 x^2 \Big)
       + \liTWmx    \Big(  - 8 + 16 x \Big)
\nonumber\\[2ex]
&&       + \liTWONmx    \Big(  - 8 + 32 x - 24 x^2 \Big)
       + \logx    \Big\{  - 8 + 28 x + 8 x^2  \zeta_2  
\nonumber\\[2ex]
&&       + \liTWmx    \Big( 24 x^2 \Big)
       + \liTWONmx    \Big( 16 x^2 \Big) \Big\}
       + \logTWx    \Big\{  - 2 + 12 x - 12 x^2 \Big\}
\nonumber\\[2ex]
&&       + \logTHx    \Big\{ {4 \over 3} x^2 \Big\}
       + \logONpx   \Big\{   - 24 x^2  \zeta_2  
       + \liTWmx    \Big(  - 48 x^2 \Big)
\nonumber\\[2ex]
&&       + \logx    \Big(  - 8 + 16 x \Big)
       + \logTWx    \Big( 20 x^2 \Big) \Big\}
       + \logTWOpx   \logx    \Big\{  - 24 x^2 \Big\}
       \Bigg]
\end{eqnarray}
For charged scalar production, 
only part of $t$-channel processes (C or D in fig.(\ref{fig9}))
interfere with $s$-channel processes in of (A and B in fig.(\ref{fig8})),
that is, 
either ($\Delta^{(2),A\overline C} _{q\overline q},\Delta^{(2),B\overline C} _{q\overline q}$)
or ($\Delta^{(2),A\overline D} _{q\overline q},\Delta^{(2),B\overline D} _{q\overline q}$)
will contribute.

To order $a_s^2$, quark (anti-quark) gluon initiated processes also receive contributions  
namely from one loop contributions to $q_i (\overline q_i) + g \rightarrow q_i(\overline q_j) +\phi$
and processes involving an additional gluon in the final state,
$q_i (\overline q_i) + g \rightarrow q_i(\overline q_j) + g+\phi$.  This contribution  
is found to be $\Delta^{(2)} _{qg} = \Delta^{(2)} _{\overline qg} = 
\Delta^{(2),C_A} _{q(\overline q)g} + \Delta^{(2),C_F} _{q(\overline q) g} $,  
where 
\begin{eqnarray}
   \Delta^{(2),C_A} _{q(\overline q)g} &=&
        C_A   T_f    
       \Bigg[
           {395 \over 9} - {208 \over 27 x} - {32 \over 3 x} \zeta_2  - 4 
         \zeta_3  + 16  \zeta_2  - {866 \over 9} x - 16 x  \zeta_3  - 80 x  \zeta_2 
\nonumber\\[2ex]
&&          + {1513 \over 27} x^2 - 8 x^2 \zeta_3  + {320 \over 3} x^2  \zeta_2  
       + \liTHpm    \Big( 16 + 32 x + 32 x^2 \Big)
\nonumber\\[2ex]
&&       + \liTHmp    \Big(  - 16 - 32 x - 32 x^2 \Big)
       + \sONTWOmx    \Big( 32 + 96 x + 32 x^2 \Big)
\nonumber\\[2ex]
&&       + \liTHmx    \Big(  - 8 - 16 x - 16 x^2 \Big)
       + \liTHOmx    \Big(  - 56 - 144 x - 48 x^2 \Big)
\nonumber\\[2ex]
&&       + \liTWmx    \Big( 12 + 32 x + 20 x^2 \Big)
       + \liTWONmx    \Bigg(  - 16 + {64 \over 3 x} + 64 x 
\nonumber\\[2ex]
&&       + {212 \over 3} x^2 \Bigg)
       + \logx    \Bigg\{ {68 \over 3} - {44 \over 3} x - 80 x  \zeta_2  - {1130 \over 9} x^2
          + 32 x^2 \zeta_2  
\nonumber\\[2ex]
&&       + \liTWmx    \Big( 16 + 32 x 
       + 32 x^2 \Big)
       + \liTWONmx    \Big(  - 16 + 32 x - 16 x^2 \Big)\Bigg\}
\nonumber\\[2ex]
&&       + \logTWx    \Bigg\{  - 1 + 36 x - {292 \over 3} x^2 \Bigg\}
       + \logTHx    \Bigg\{ {10 \over 3} + {28 \over 3} x \Bigg\}
       + \logONpx   
\nonumber\\[2ex]
&&     \times \Bigg\{\logx    \Big( 12 + 32 x + 20 x^2 \Big)
       + \logTWx    
        \Big( 12 + 24 x + 24 x^2 \Big)\Bigg\}
\nonumber\\[2ex]
&&       + \logONmx   \Bigg\{   - {152 \over 3} + {32 \over 9 x} - 32  \zeta_2  + {38 \over 3} 
         x + 32 x \zeta_2  + {364 \over 9} x^2 - 64 x^2  \zeta_2  
\nonumber\\[2ex]
&&       + \liTWmx    \Big(  - 16 - 32 x - 32 x^2 \Big)
       + \liTWONmx    \Big( 52 + 88 x + 40 x^2 \Big)
\nonumber\\[2ex]
&&       + \logx    \Big( 4 - 112 x + 248 x^2 \Big)
       + \logTWx    \Big(  - 12 - 56 x + 8 x^2 \Big)
\nonumber\\[2ex]
&&       + \logONpx   \logx    \Big(  - 16 - 32 x - 32 x^2 \Big) \Bigg\}
       + \logTWOmx   \Bigg\{   - 10 + {32 \over 3 x} 
\nonumber\\[2ex]
&&     + 96 x - {290 \over 3} x^2 
       + \logx    ( 4 + 88 x - 24 x^2 )\Bigg\}
       + \logTHOmx    \Bigg\{ {26 \over 3} - {52 \over 3} x 
\nonumber\\[2ex]
&&       + {52 \over 3} x^2 \Bigg\}
       + \logqmuf \Bigg\{     - {82 \over 3} + {16 \over 9 x} - 16  \zeta_2  + {16 \over 3} 
         x + 16 x \zeta_2  + {218 \over 9} x^2 
\nonumber\\[2ex]
&&       - 32 x^2  \zeta_2  
       + \liTWmx    \Big(  - 8 - 16 x - 16 x^2 \Big)
       + \liTWONmx    \Big( 24 + 48 x 
\nonumber\\[2ex]
&&       + 16 x^2 \Big)
       + \logx    \Big(  - 4 - 40 x + 112 x^2 \Big)
       + \logTWx    \Big(  - 8 - 24 x \Big)
\nonumber\\[2ex]
&&       + \logONpx   \logx    \Big(  - 8 - 16 x - 16 x^2 \Big)
       + \logONmx    \Bigg(  - 4 + {32 \over 3 x} + 88 x 
\nonumber\\[2ex]
&&      - {284 \over 3} 
         x^2 \Bigg)
       + \logONmx   \logx    \Big( 8 + 80 x - 16 x^2 \Big)
       + \logTWOmx    \Big( 12 - 24 x 
\nonumber\\[2ex]
&&     + 24 x^2 \Big)\Bigg\}
       + \logTWqmuf    \Bigg\{ 2 + {8 \over 3 x} + 16 x - {62 \over 3} x^2 
       + \logx    ( 4 + 16 x )
\nonumber\\[2ex]
&&       + \logONmx    ( 4 - 8 x + 8 x^2 ) \Bigg\}
       \Bigg]
\end{eqnarray}
and 
\begin{eqnarray}
   \Delta^{(2),C_F} _{q(\overline q)g} 
\nonumber\\[2ex]
       &=& C_F   T_f    
       \Bigg[
           - {129 \over 2} + 76 \zeta_3  + 20  \zeta_2  + 329 x - 152 
         x \zeta_3  - 64 x  \zeta_2  - {549 \over 2} x^2 
\nonumber\\[2ex]
&&       + 200 x^2  \zeta_3  + 24 x^2
          \zeta_2  
       + \sONTWOmx    \Big(  - 28 + 56 x - 136 x^2 \Big)
\nonumber\\[2ex]
&&       + \liTHmx    \Big( 64 x^2 \Big)
       + \liTHOmx    \Big(  - 12 + 24 x + 72 x^2 \Big)
\nonumber\\[2ex]
&&       + \liTWmx    \Big(  - 8 - 32 x - 24 x^2 \Big)
       + \liTWONmx    \Big(  - 26 - 32 x + 56 x^2 \Big)
\nonumber\\[2ex]
&&       + \logx   \Big\{   - 35 + 16  \zeta_2  + 301 x - 32 x  \zeta_2 
          - 214 x^2 + 96 x^2 \zeta_2  
\nonumber\\[2ex]
&&       + \liTWmx    \Big(  - 32 x^2 \Big)
       + \liTWONmx    \Big( 4 - 8 x \Big) \Big\}
       + \logTWx    \Bigg\{  - {19 \over 2} + 70 x 
\nonumber\\[2ex]
&&       - 38 x^2 \Bigg\}
       + \logTHx    \Bigg\{  - 3 + 6 x - {52 \over 3} x^2 \Bigg\}
       + \logONpx   \logx    \Big(  - 8 - 32 x 
\nonumber\\[2ex]
&&    - 24 x^2 \Big)
       + \logONmx  \Bigg\{   128 - 16  \zeta_2  - 394 x + 32 x  \zeta_2 
          + 272 x^2 - 32 x^2 \zeta_2  
\nonumber\\[2ex]
&&       + \liTWONmx    \Big(  - 4 + 8 x - 104 x^2 \Big)
       + \logx    \Big( 56 - 256 x + 216 x^2 \Big)
\nonumber\\[2ex]
&&       + \logTWx    \Big( 24 - 48 x + 80 x^2 \Big) \Bigg\}
       + \logTWOmx   \Bigg\{   - 66 + 192 x - 138 x^2 
\nonumber\\[2ex]
&&       + \logx    \Big(  - 42 + 84 x - 132 x^2 \Big) \Bigg\}
       + \logTHOmx    \Bigg\{ {70 \over 3} - {140 \over 3} x + {140 \over 3} x^2 \Bigg\} 
\nonumber\\[2ex]
&&       + \logqmuf    \Bigg\{ 54 - 8  \zeta_2  - 172 x + 16 x  \zeta_2 
          + 120 x^2 - 16 x^2 \zeta_2  
\nonumber\\[2ex]
&&       + \liTWONmx    \Big(  - 48 x^2 \Big)
       + \logx    \Big( 22 - 128 x + 116 x^2 \Big)
       + \logTWx    \Big( 8 - 16 x 
\nonumber\\[2ex]
&&       + 32 x^2 \Big)
       + \logONmx    \Big(  - 72 + 200 x - 140 x^2 \Big)
       + \logONmx   \logx    \Big(  - 40 
\nonumber\\[2ex]
&&        + 80 x 
       - 128 x^2 \Big)
       + \logTWOmx    \Big( 36 - 72 x + 72 x^2 \Big) \Bigg\}
       + \logTWqmuf    \Bigg\{  - 15 
\nonumber\\[2ex]
&&       + 36 x - 24 x^2 
       + \logx    \Big(  - 6 + 12 x - 24 x^2 \Big)
       + \logONmx   
\nonumber\\[2ex] 
&&     \times \Big( 12 - 24 x + 24 x^2 \Big) \Bigg\}
       \Bigg]
\end{eqnarray}
We now present the contributions resulting from pure $t$-channel processes
where the final state quarks (anti-quarks) are non-identical (C and D in fig.(\ref{fig9})).  
The individual $t$-channel results are found to be identical and are given by  
\begin{eqnarray}
   \Delta^{(2),C \overline C} _{q\overline q} &=&
   \Delta^{(2),D \overline D} _{q\overline q} =
   \Delta^{(2),C \overline C} _{q q} =
   \Delta^{(2),D \overline D} _{q q} =
   \Delta^{(2),C \overline C} _{\overline q \overline q} =
   \Delta^{(2),D \overline D} _{\overline q \overline q} 
\nonumber\\[2ex]        
&=&     C_F   T_f    
       \Bigg[
          {305 \over 9} - {208 \over 27 x} - {32 \over 3 x} \zeta_2  - 8 
         \zeta_2  - {470 \over 9} x + 8 x  \zeta_2  + {703 \over 27} x^2 + {32 \over 3} x^2  \zeta_2 
\nonumber\\[2ex]
&&       + \sONTWOmx    \Big( 16 + 16 x \Big)
       + \liTHOmx    \Big(  - 32 - 32 x \Big)
       + \liTWONmx 
\nonumber\\[2ex]
&&       \times   \Bigg(  - 4 + {64 \over 3 x} + 28 x + {32 \over 3} x^2 \Bigg)
       + \logx    \Bigg\{ {62 \over 3} - 16  \zeta_2  - {176 \over 3} x - 16 x 
         \zeta_2  
\nonumber\\[2ex]
&&     + {40 \over 9} x^2 
       +  \liTWONmx    \Big(  - 8 - 8 x \Big) \Bigg\}
       + \logTWx    \Bigg\{  - 1 - 5 x - {40 \over 3} x^2 \Bigg\}
\nonumber\\[2ex]
&&       + \logTHx    \Bigg\{ {10 \over 3} + {10 \over 3} x \Bigg\}
       + \logONmx    \Bigg\{  - {152 \over 3} + {32 \over 9 x} + {200 \over 3} x - {176 \over 9} x^2
\nonumber\\[2ex]
&&       + \liTWONmx    \Big( 32 + 32 x \Big)
       + \logx    \Big(  - 8 + 32 x + 32 x^2 \Big)
\nonumber\\[2ex]
&&       + \logTWx    \Big(  - 16 - 16 x \Big) \Bigg\}
       + \logTWOmx    \Bigg\{ 8 + {32 \over 3 x} - 8 x - {32 \over 3} x^2 
\nonumber\\[2ex]
&&       + \logx    \Big( 16 + 16 x \Big)\Bigg\}
       + \logqmuf    \Bigg\{  - {76 \over 3} + {16 \over 9 x} + {100 \over 3} x - {88 \over 9} x^2 
\nonumber\\[2ex]
&&       + \liTWONmx    \Big( 16 + 16 x \Big)
       + \logx    \Big(  - 4 + 16 x + 16 x^2 \Big)
\nonumber\\[2ex]
&&       + \logTWx    \Big(  - 8 - 8 x \Big)
       + \logONmx    \Bigg( 8 + {32 \over 3 x} - 8 x - {32 \over 3} x^2 \Bigg)
\nonumber\\[2ex]
&&       + \logONmx   \logx    \Big( 16 + 16 x \Big) \Bigg\}
       + \logTWqmuf    \Bigg\{ 2 + {8 \over 3 x} - 2 x - {8 \over 3} x^2 
\nonumber\\[2ex]
&&       +  \logx    \Big( 4 + 4 x \Big) \Bigg\}
       \Bigg]
\end{eqnarray}
If the final state quarks (anti-quarks) are identical, we find in addition
to above mentioned $t$-channel processes, we have to include similar $t$ processes
with final state quarks (anti-quarks) interchanged (fig.(\ref{fig10})).  Hence we have
contributions from the additional $t$ processes (fig.(\ref{fig10})) and their interferences
with the already existing $t$-channel processes (fig.(\ref{fig9})).  
The former gives results identical to $\Delta^{(2),C\overline C}$,  
$\Delta^{(2),D\overline D}$  and  $\Delta^{(2),C\overline D}$.  
The interference contributions are found to be
\begin{eqnarray}
   \Delta^{(2),C \overline E} _{q q} &=&
   \Delta^{(2),D \overline F} _{q q} =
   \Delta^{(2),C \overline E} _{\overline q \overline q} =
   \Delta^{(2),D \overline F} _{\overline q \overline q} 
\nonumber\\[2ex]        
&=&      \Bigg(C_F-{C_A \over 2}\Bigg)   C_F    
        \Bigg[
           - 34 - {8 \over 1+x }  \zeta_3  + 4  \zeta_3  + 4 
         \zeta_2  + 34 x - 4 x  \zeta_3  
\nonumber\\[2ex]
&&       + \liTHpm    \Bigg(  - 32 + {64 \over  1+x } + 32 x \Bigg)
       + \liTHmp    \Bigg( 32 
\nonumber\\[2ex]
&&       - {64 \over  1+x } - 32 x \Bigg)
       + \sONTWmx    \Bigg( 40 - {32 \over  1+x } - 40 x \Bigg)
       + \sONTWOmx  
\nonumber\\[2ex]
&&     \times  \Bigg(  - 32 + {64 \over  1+x } + 32 x \Bigg)
       + \liTHmx    \Bigg( 12 - {16 \over  1+x } - 12 x \Bigg)
\nonumber\\[2ex]
&&       + \liTHOmx    \Bigg( 32 - {64 \over  1+x } - 32 x \Bigg)
       + \liTWmx    \Big( 8 \Big)
       + \liTWONmx 
\nonumber\\[2ex]
&&     \times   \Big( 24 + 8 x \Big)
       + \logx    \Bigg\{  - 18 + {24 \over  1+x }  \zeta_2  - 16 
         \zeta_2  + 14 x + 16 x  \zeta_2  
\nonumber\\[2ex]
&&       + \liTWmx    (  - 44 + {64 \over  1+x } + 44 x )
       + \liTWONmx    (  - 24 + {48 \over  1+x } + 24 x ) \Bigg\}
\nonumber\\[2ex]
&&       + \logTWx    \Big\{  - 4 - 10 x \Big\}
       + \logTHx    \Bigg\{ {10 \over 3} - {16 \over 3  1+x } - {10 \over 3} x \Bigg\}
       + \logONpx 
\nonumber\\[2ex]
&&     \times  \Bigg\{  - {16 \over  1+x }  \zeta_2  + 20  \zeta_2 
          - 20 x \zeta_2  
       + \liTWmx    \Bigg( 40 - {32 \over  1+x } - 40 x \Bigg)
\nonumber\\[2ex]
&&       + \logx    \Big( 8 \Big)
       + \logTWx    \Bigg(  - 38 + {56 \over  1+x } + 38 x \Bigg) \Bigg\}
       + \logTWOpx   \logx  
\nonumber\\[2ex]
&&      \times \Bigg\{ 20 - {16 \over  1+x } - 20 x \Bigg\}
       + \logONmx    \Bigg\{ 32 - {32 \over  1+x }  \zeta_2  + 16 
         \zeta_2  - 32 x 
\nonumber\\[2ex]
&&     - 16 x  \zeta_2  
       + \liTWmx    \Bigg( 32 - {64 \over  1+x } - 32 x \Bigg)
       + \logx    \Big( 16 + 16 x \Big)
\nonumber\\[2ex]
&&       + \logTWx    \Bigg(  - 8 + {16 \over  1+x } + 8 x \Bigg)
       + \logONpx   \logx    \Bigg( 32 - {64 \over  1+x } - 32
          x \Bigg) \Bigg\}
\nonumber\\[2ex]
&&       + \logqmuf    \Bigg\{ 16 - {16 \over  1+x }  \zeta_2  + 8 
         \zeta_2  - 16 x - 8 x  \zeta_2  
       + \liTWmx    \Bigg( 16 
\nonumber\\[2ex]
&&     - {32 \over  1+x } - 16 x \Bigg)
       + \logx    \Big( 8 + 8 x \Big)
       + \logTWx    \Bigg(  - 4 + {8 \over  1+x } + 4 x \Bigg)
\nonumber\\[2ex]
&&       + \logONpx   \logx    \Bigg( 16 - {32 \over  1+x } - 
         16 x \Bigg)\Bigg\}
        \Bigg]
\end{eqnarray}
and 
\begin{eqnarray}
\\[2ex]
   \Delta^{(2),C\overline F} _{qq} &=&
   \Delta^{(2),D \overline E} _{ qq} =
   \Delta^{(2),C \overline F} _{\overline q\overline q} =
   \Delta^{(2),D \overline E} _{\overline q\overline q} 
\nonumber\\[2ex]        
&=&          \Bigg(C_F-{C_A \over 2}\Bigg)   C_F    
        \Bigg[
           - 23 + 36 x - 13 x^2 
       + \sONTWOmx    \Big( 4 - 8 x - 8 x^2 \Big)
\nonumber\\[2ex]
&&       + \liTHOmx    \Big( 4 - 8 x + 8 x^2 \Big)
       + \liTWONmx    \Big( 4 + 12 x - 12 x^2 \Big)
\nonumber\\[2ex]
&&       + \logx    \Big\{ 2 - 4 x 
       + \liTWONmx    \Big(  - 8 x^2 \Big)\Big\}
       + \logTWx    \Big( 2 + 6 x - 6 x^2 \Big)
\nonumber\\[2ex]
&&       + \logTHx    \Big(  - {4 \over 3 }x^2 \Big)
      \Bigg]
\end{eqnarray}
Finally we compute gluon gluon initiated processes that contribute to order $a_s^2$. 
Writing $\Delta^{(2)}_{gg}=\Delta^{(2),C_A}_{gg}+\Delta^{(2),C_F}_{gg}$, we find
\begin{eqnarray}
   \Delta^{(2),C_A} _{gg} &=&
        {N^2 \over N^2-1 }   
         \Bigg[
          - {1 \over 3} + 8 \zeta_3  - {248 \over 3} x + 16 x  \zeta_3  - 
         4 x \zeta_2  + 83 x^2 + 16 x^2  \zeta_3  + {8 \over 3} x^2  \zeta_2  
\nonumber\\[2ex]
&&       + \sONTWmx    \Big( 8 + 16 x + 16 x^2 \Big)
       + \sONTWOmx    \Big(  - 4 + 8 x - 8 x^2 \Big)
\nonumber\\[2ex]
&&       + \liTHmx    \Big( 12 + 24 x + 24 x^2 \Big)
       + \liTWmx    \Bigg(  - 8 x + {16 \over 3} x^2 \Bigg)
\nonumber\\[2ex]
&&       + \logx    \Bigg\{ {4 \over 3} - {80 \over 3} x - 58 x^2 
       + \liTWmx    \Big(  - 12 - 24 x - 24 x^2 \Big)\Bigg\}
\nonumber\\[2ex]
&&       + \logTWx    \Bigg\{ 2 x + {50 \over 3} x^2 \Bigg\}
       + \logONpx    \Big\{ 4  \zeta_2  + 8 x  \zeta_2  + 8 x^2 
         \zeta_2  
\nonumber\\[2ex]
&&       + \liTWmx    \Big( 8 + 16 x + 16 x^2 \Big)
       + \logx    \Bigg(  - 8 x + {16 \over 3} x^2 \Bigg)
       + \logTWx    
\nonumber\\[2ex]
&&     \times \Bigg(  - 6 - 12 x - 12 x^2 \Bigg) \Bigg\}
       + \logTWOpx   \logx    \Bigg\{ 4 + 8 x + 8 x^2 \Bigg\}
        \Bigg]
\end{eqnarray}
and
\begin{eqnarray}
   \Delta^{(2),C_F} _{gg} &=&
        \sONTWmx    \Big(  - 8 - 16 x - 16 x^2 \Big)
       + \sONTWOmx    \Big(  - 12 - 72 x - 56 x^2 \Big)
\nonumber\\[2ex]
&&       + \liTHmx    \Big(  - 12 - 24 x + 8 x^2 \Big)
       + \liTHOmx    \Big( 16 + 64 x + 64 x^2 \Big)
\nonumber\\[2ex]
&&       + \liTWmx    \Big( 8 x \Big)
       + \liTWONmx    \Big(  - 4 + 16 x + 56 x^2 \Big)
       + \logx  
\nonumber\\[2ex]
&&   \times  \Bigg\{  - 15 + 8  \zeta_2  - 48 x + 32 x  \zeta_2  + 121 x^2 + 
         40 x^2 \zeta_2  
       + \liTWmx    \Big( 12 
\nonumber\\[2ex]
&&    + 24 x + 8 x^2 \Big)
       + \liTWONmx    \Big(  - 4 - 16 x - 16 x^2 \Big)\Bigg\}
       + \logTWx    \Big\{  - 4 
\nonumber\\[2ex]
&&    - 30 x - 8 x^2 \Big\}
       + \logTHx    \Bigg\{  - {2 \over 3} - {8 \over 3} x - {16 \over 3} x^2 \Bigg\}
       + \logONpx  \Bigg\{    - 4  \zeta_2  
\nonumber\\[2ex]
&&     - 8 x  \zeta_2  - 8 x^2  \zeta_2  
       + \liTWmx    \Big(  - 8 - 16 x - 16 x^2 \Big)
       + \logx    \Big( 8 x \Big)
\nonumber\\[2ex]
&&       + \logTWx    \Big( 6 + 12 x + 12 x^2 \Big) \Bigg\}
       + \logTWOpx   \logx    \Big\{  - 4 - 8 x - 8 x^2 \Big\}
\nonumber\\[2ex]
&&       + \logONmx   \Big\{  46 + 104 x - 150 x^2 
       + \liTWONmx    \Big(  - 16 - 64 x - 64 x^2 \Big)
\nonumber\\[2ex]
&&       + \logx    \Big( 20 + 64 x - 16 x^2 \Big)
       + \logTWx    \Big( 4 + 16 x + 16 x^2 \Big) \Big\}
\nonumber\\[2ex]
&&       + \logTWOmx\Big\{      - 16 - 32 x + 48 x^2 
       + \logx    \Big(  - 8 - 32 x - 32 x^2 \Big) \Big\}
\nonumber\\[2ex]
&&       + \logqmuf \Bigg\{    23 + 52 x - 75 x^2 
       + \liTWONmx    \Big(  - 8 - 32 x - 32 x^2 \Big)
\nonumber\\[2ex]
&&       + \logx    \Big( 10 + 32 x - 8 x^2 \Big)
       + \logTWx    \Big( 2 + 8 x + 8 x^2 \Big)
       + \logONmx 
\nonumber\\[2ex]
&&       \times    \Big(  - 16 - 32 x + 48 x^2 \Big)
       + \logONmx   \logx    \Big(  - 8 - 32 x - 32 x^2 \Big) \Bigg\}
\nonumber\\[2ex]
&&       + \logTWqmuf  \Bigg\{    - 4 - 8 x + 12 x^2 
       + \logx    \Big(  - 2 - 8 x - 8 x^2 \Big) \Bigg\}
\nonumber\\[2ex]
&&       - 20 - 8 \zeta_3  + 16  \zeta_2  - 98 x - 16 x  \zeta_3  + 36 x 
         \zeta_2  + 118 x^2 + 8 x^2  \zeta_3  - 48 x^2  \zeta_2 
\end{eqnarray}

\newcommand{\plb}[3]{{Phys.Lett.} {\bf B#1} (#3) #2}                  %
\newcommand{\prl}[3]{Phys.Rev.Lett. {\bf #1} (#3) #2}        %
\newcommand{\rmp}[3]{Rev.Mod.Phys. {\bf #1} (#3) #2}             %
\newcommand{\prep}[3]{Phys.Rept. {\bf #1} (#3) #2}                     %
\newcommand{\rpp}[3]{Rept.Prog.Phys. {\bf #1} (#3) #2}             %
\newcommand{\prd}[3]{{Phys.Rev.} {\bf D#1} (#3) #2}                    %
\newcommand{\np}[3]{Nucl.Phys. {\bf B#1} (#3) #2}                     %
\newcommand{\npbps}[3]{Nucl.Phys. B (Proc. Suppl.)
           {\bf #1} (#3) #2}                                           %
\newcommand{\sci}[3]{Science {\bf #1} (#3) #2}                 %
\newcommand{\zp}[3]{Z.Phys. C{\bf#1} (#3) #2}                 %
\newcommand{\mpla}[3]{Mod.Phys.Lett. {\bf A#1} (#3) #2}             %
\newcommand{\astropp}[3]{Astropart.Phys. {\bf #1} (#3) #2}            %
\newcommand{\ib}[3]{{\em ibid.\/} {\bf #1} (#3) #2}                    %
\newcommand{\nat}[3]{Nature (London) {\bf #1} (#3) #2}         %
\newcommand{\nuovocim}[3]{Nuovo Cim. {\bf #1} (#3) #2}         %
\newcommand{\yadfiz}[4]{Yad.Fiz. {\bf #1} (#3) #2 [English            %
        transl.: Sov.J.Nucl. Phys. {\bf #1} #3 (#4)]}               %
\newcommand{\philt}[3]{Phil.Trans.Roy.Soc. London A {\bf #1} #2
        (#3)}                                                          %
\newcommand{\hepph}[1]{(electronic archive:     hep--ph/#1)}           %
\newcommand{\hepex}[1]{(electronic archive:     hep--ex/#1)}           %
\newcommand{\astro}[1]{(electronic archive:     astro--ph/#1)}         %

\begin{figure}
\begin{center}
  \begin{picture}(82,98) (76,-31)
    \SetScale{0.7}
    \SetWidth{1.0}
    \SetColor{Black}
    \Line(97,66)(198,17)
    \Line(97,-30)(197,17)
    \Line[dash,dashsize=5](197,17)(277,17)
  \end{picture}
\caption[]{Subprocess $q_i+\bar q_j
\rightarrow \phi $. }
\label{fig1}
\end{center}
\end{figure}
\begin{figure}
\begin{center}
  \begin{picture}(351,79) (79,-24)
    \SetScale{0.9}
    \SetWidth{1.0}
    \SetColor{Black}
    \Line(82,48)(146,17)
    \Line(145,16)(80,-16)
    \Gluon(101,38)(102,-5){5}{3}
    \Line[dash,dashsize=5](146,17)(199,17)
    \Line(217,47)(281,16)
    \Line(281,15)(216,-17)
    \Line[dash,dashsize=5](282,15)(335,15)
    \Line(354,47)(418,16)
    \Line(417,16)(352,-16)
    \Line[dash,dashsize=5](419,16)(472,16)
    \GluonArc[clock](252.063,31.664)(17.268,158.475,-30.117){5}{5}
    \GluonArc(387.128,1.244)(19.009,-154.297,27.427){5}{5}
  \end{picture}
\caption[]{Subprocess $q_i+\bar q_j
\rightarrow \phi $. }
\label{fig2}
\end{center}
\end{figure}
\begin{figure}
\begin{center}
  \begin{picture}(263,61) (31,-25)
    \SetScale{0.9}
    \SetWidth{1.0}
    \SetColor{Black}
    \Line(33,29)(96,29)
    \Line(97,30)(97,-18)
    \Line(97,-19)(32,-19)
    \Gluon(96,30)(147,29){5}{4}
    \Line[dash,dashsize=5](97,-18)(146,-18)
    \Line(177,29)(240,29)
    \Line(241,-17)(176,-17)
    \Line(241,30)(241,-18)
    \Line[dash,dashsize=5](241,29)(290,29)
    \Gluon(242,-18)(293,-19){5}{4}
  \end{picture}
\caption[]{Subprocess $q_i+\bar q_j
\rightarrow \phi + g  $. }
\label{fig3}
\end{center}
\end{figure}
\begin{figure}
\begin{center}
  \begin{picture}(227,59) (48,-42)
    \SetScale{0.9}
    \SetWidth{1.0}
    \SetColor{Black}
    \Line(49,12)(98,12)
    \Line(96,12)(96,-37)
    \Line(145,12)(97,12)
    \Gluon(49,-36)(96,-36){5}{4}
    \Line[dash,dashsize=5](96,-36)(144,-36)
    \Line(274,13)(226,13)
    \Line(176,-37)(225,-37)
    \Gluon(178,11)(225,11){5}{4}
    \Line(226,13)(226,-36)
    \Line[dash,dashsize=5](226,-37)(274,-37)
  \end{picture}
\caption[]{Subprocess $q_i+g
\rightarrow \phi + q_j $. }
\label{fig4}
\end{center}
\end{figure}
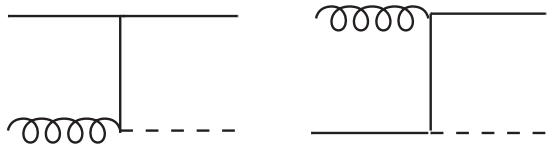
\begin{figure}
\begin{center}
  \begin{picture}(242,452) (78,-76)
    \SetScale{0.7}
    \SetWidth{1.0}
    \SetColor{Black}
    \Line(85,375)(160,311)
    \Line(160,310)(85,243)
    \Line[dash,dashsize=5](164,310)(210,310)
    \Gluon(128,337)(129,283){5}{4}
    \Gluon(105,356)(106,263){5}{7}
    \Line(241,372)(302,308)
    \Line(300,308)(239,241)
    \Line[dash,dashsize=5](302,308)(352,308)
    \Gluon(253,357)(272,277){5}{7}
    \Gluon(98,198)(99,167){5}{3}
    \Gluon(100,127)(100,97){5}{2}
    \Arc(262,145)(17,90,450)
    \Gluon(260,193)(260,162){5}{3}
    \Gluon(260,130)(258,99){5}{2}
    \Gluon(397,201)(395,93){5}{9}
    \Line(385,374)(464,310)
    \Line(464,309)(384,242)
    \Line[dash,dashsize=5](464,309)(519,309)
    \Arc[dash,dashsize=2](405,311)(15.133,98,458)
    \Gluon(405,358)(403,326){5}{3}
    \Gluon(404,297)(404,259){5}{3}
    \Line(162,-12)(79,-75)
    \Line(79,52)(162,-12)
    \Line[dash,dashsize=5](163,-12)(210,-12)
    \Gluon(119,20)(120,-44){5}{5}
    \GluonArc[clock](123.069,27.8)(22.266,161.132,-57.6){5}{7}
    \Line(322,-13)(240,-75)
    \Line(241,52)(323,-12)
    \Line[dash,dashsize=5](322,-12)(377,-12)
    \Gluon(252,44)(252,-65){5}{9}
    \GluonArc[clock](286.451,18.713)(23.556,145.663,-41.841){5}{6}
    \Line(81,212)(160,148)
    \Line(162,148)(81,81)
    \Line[dash,dashsize=5](162,147)(210,147)
    \GluonArc(100,146)(14.56,-74,286){5}{7}
    \Line(241,212)(312,148)
    \Line(310,147)(239,80)
    \Line[dash,dashsize=5](313,148)(360,148)
    \Line(386,212)(453,148)
    \Line(452,147)(384,81)
    \Line[dash,dashsize=5](455,148)(517,148)
    \Gluon(401,162)(437,132){5}{4}
    \Gluon(268,317)(278,334){5}{1}
    \Gluon(256,261)(264,306){5}{4}
    \Line(378,51)(415,17)
    \Line(415,16)(415,-34)
    \Line(415,-34)(373,-69)
    \Gluon(415,16)(451,2){5}{3}
    \Gluon(416,-34)(453,-26){5}{3}
    \Line(452,6)(452,-25)
    \Line(453,-24)(480,-6)
    \Line(480,-6)(452,6)
    \Line[dash,dashsize=5](482,-6)(521,-6)
  \end{picture}
\caption[]{Subprocess $q_i+\bar q_j
\rightarrow \phi $. }
\label{fig5}
\end{center}
\end{figure}
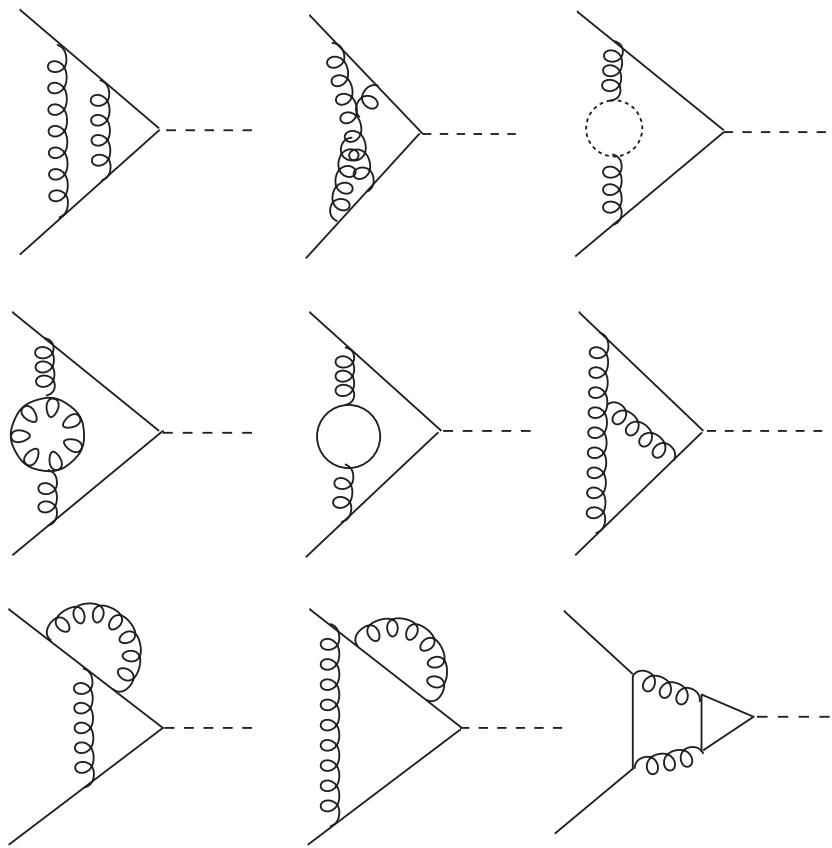

\begin{figure}
\begin{center}
  \begin{picture}(200,200) (63,-91)
    \SetScale{0.7}
    \SetWidth{1.0}
    \SetColor{Black}
    \Line(64,248)(144,248)
    \Line(144,248)(144,166)
    \Line(144,166)(64,166)
    \Gluon(95,247)(94,167){5}{6}
    \Line[dash,dashsize=5](144,248)(194,248)
    \Gluon(145,167)(192,167){5}{4}
    \Line(65,135)(152,98)
    \Line(151,97)(64,52)
    \Gluon(152,97)(209,95){5}{4}
    \Line[dash,dashsize=5](84,62)(150,62)
    \Gluon(102,119)(101,72){5}{4}
    \Line(302,92)(227,38)
    \Line(228,133)(303,92)
    \Line[dash,dashsize=5](303,92)(353,92)
    \Gluon(264,113)(263,65){5}{4}
    \Gluon(248,52)(321,39){5}{6}
    \Line(386,137)(461,96)
    \Line(461,96)(386,42)
    \Line[dash,dashsize=5](462,96)(512,96)
    \Gluon(379,90)(404,91){5}{2}
    \Gluon(404,93)(420,117){5}{2}
    \Gluon(404,92)(428,72){5}{2}
    \Line(288,167)(208,167)
    \Gluon(304,246)(351,246){5}{4}
    \Line(304,167)(224,167)
    \Line(304,248)(304,166)
    \Line(225,248)(305,248)
    \Line[dash,dashsize=5](305,167)(355,167)
    \GluonArc(295.125,208)(24.311,71.099,288.901){5}{7}
    \Line(449,249)(449,167)
    \Line(368,249)(448,249)
    \Line(449,167)(369,167)
    \Gluon(449,166)(511,166){5}{5}
    \Line[dash,dashsize=5](452,249)(502,249)
    \Line[dash,dashsize=5](450,249)(511,249)
    \Gluon(449,215)(476,170){5}{4}
    \Line(66,7)(111,-37)
    \Line(111,-37)(65,-87)
    \Gluon(110,-39)(180,-39){5}{6}
    \Line(179,-37)(220,5)
    \Line(220,6)(220,-89)
    \Line(179,-37)(221,-90)
    \Gluon(219,3)(275,2){5}{4}
    \Line[dash,dashsize=5](222,-90)(275,-90)
  \end{picture}
\caption[]{Subprocess $q_i+\bar q_j
\rightarrow \phi + g $. }
\label{fig6}
\end{center}
\end{figure}
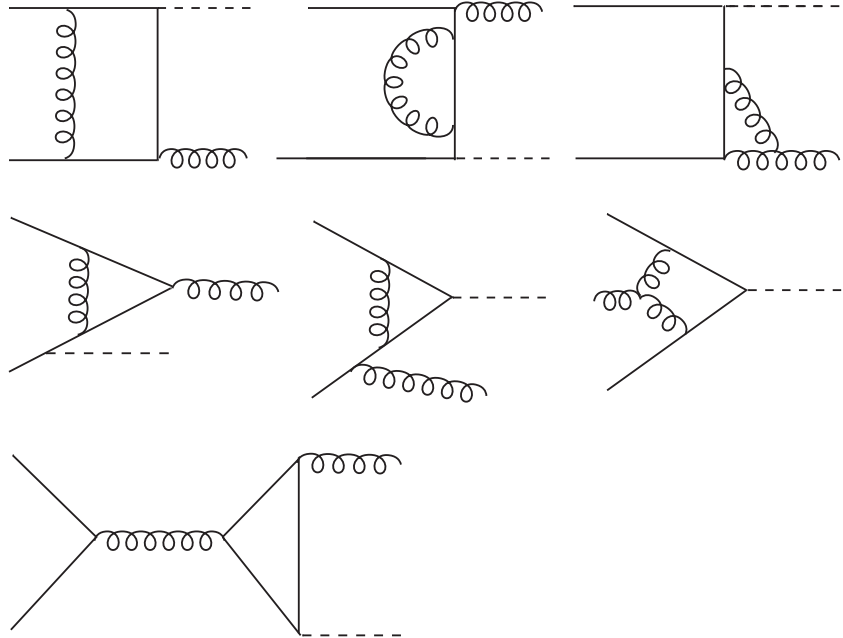

\begin{figure}
\begin{center}
  \begin{picture}(197,124) (62,-76)
    \SetScale{0.7}
    \SetWidth{1.0}
    \SetColor{Black}
    \Line(65,145)(114,145)
    \Line(256,145)(256,80)
    \Line(114,80)(63,80)
    \Line[dash,dashsize=5](114,80)(170,80)
    \Gluon(113,142)(170,142){5}{6}
    \Gluon(113,112)(168,112){5}{5}
    \Line(256,80)(205,80)
    \Line(207,145)(256,145)
    \Line(114,17)(114,-48)
    \Line(401,81)(350,81)
    \Line(353,145)(402,145)
    \Line(402,146)(402,81)
    \Gluon(402,112)(457,112){5}{5}
    \Gluon(403,81)(458,81){5}{5}
    \Line[dash,dashsize=5](403,145)(459,145)
    \Gluon(256,141)(311,141){5}{5}
    \Gluon(257,80)(312,80){5}{5}
    \Line[dash,dashsize=5](257,113)(313,113)
    \Line(65,16)(114,16)
    \Line(115,-47)(64,-47)
    \Line(113,146)(113,81)
    \Line[dash,dashsize=5](115,-47)(171,-47)
    \Line(226,17)(275,17)
    \Line(275,-47)(224,-47)
    \Line[dash,dashsize=5](277,17)(333,17)
    \Line(276,17)(276,-48)
    \Gluon(114,16)(138,15){5}{2}
    \Gluon(139,17)(175,33){5}{4}
    \Gluon(139,18)(172,-13){5}{4}
    \Gluon(277,-47)(301,-48){5}{2}
    \Gluon(300,-46)(336,-30){5}{4}
    \Gluon(299,-44)(332,-75){5}{4}
  \end{picture}
\caption[]{Subprocess $q_i+\bar q_j
\rightarrow \phi + g +g$. }
\label{fig7}
\end{center}
\end{figure}
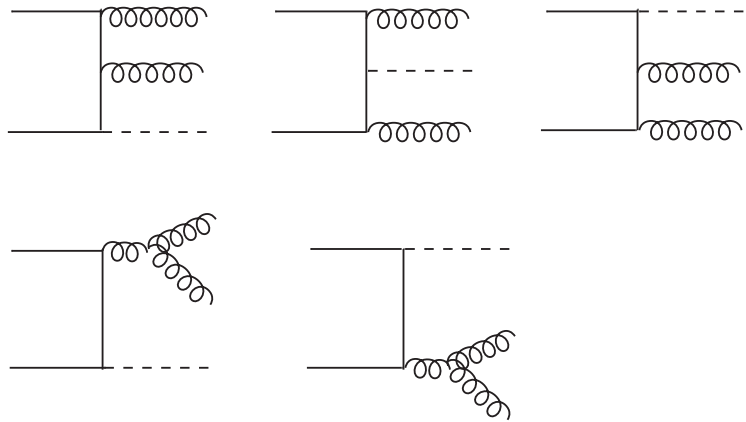

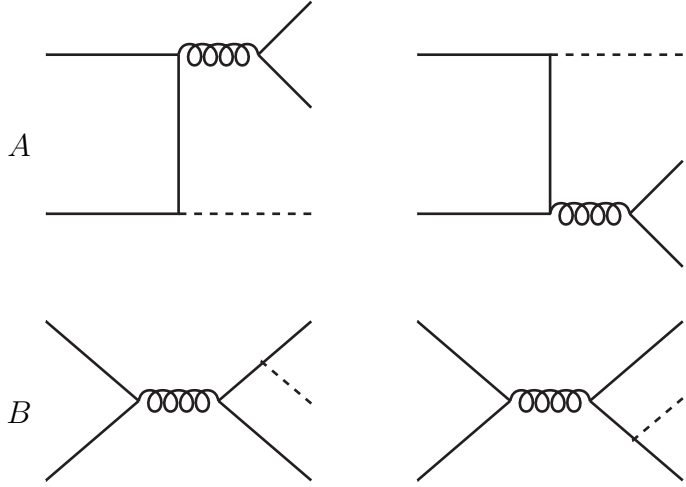
\begin{figure}
\begin{center}
\begin{picture}(260,95)(0,0)
    \SetWidth{1.0}
    \SetColor{Black}
\Text(0,50)[t]{$A$}
\Line(10,20)(60,20)
\Line(10,80)(60,80)
\Line(60,20)(60,80)
\Gluon(60,80)(90,80){4}{4}
\Line(90,80)(110,100)
\Line(90,80)(110,60)
\DashLine(60,20)(110,20){3}
\Line(150,20)(200,20)
\Line(150,80)(200,80)
\Line(200,20)(200,80)
\Gluon(200,20)(230,20){4}{4}
\Line(230,20)(250,0)
\Line(230,20)(250,40)
\DashLine(200,80)(250,80){3}

\end{picture}
\begin{picture}(260,100)(0,0)
    \SetWidth{1.0}
    \SetColor{Black}

\Text(0,50)[t]{$B$}

\Line(10,20)(45,50)
\Line(10,80)(45,50)
\Gluon(45,50)(75,50){4}{4}
\Line(75,50)(110,80)
\Line(75,50)(110,20)
\DashLine(91,65)(110,49){3}

\Line(150,20)(185,50)
\Line(150,80)(185,50)
\Gluon(185,50)(215,50){4}{4}
\Line(215,50)(250,20)
\Line(215,50)(250,80)
\DashLine(231,35)(250,51){3}
\end{picture}
\caption[]{$s$ channel annihilation graphs contributing to the subprocess $q_i+\bar q_j
\rightarrow \phi + q_k+\bar q_l$. }
\label{fig8}
\end{center}
\end{figure}
\begin{figure}
\begin{center}
 \begin{picture}(260,90)(0,0)
    \SetWidth{1.0}
    \SetColor{Black}

\Text(0,50)[t]{$C$}

\Line(10,20)(110,20)
\Line(10,70)(110,70)
\Gluon(60,20)(60,70){4}{6}
\DashLine(35,70)(70,90){3}

\Line(150,20)(250,20)
\Line(150,70)(250,70)
\Gluon(200,20)(200,70){4}{5}
\DashLine(225,70)(260,90){3}

\Text(0,70)[t]{$1$}
\Text(0,20)[t]{$2$}
\Text(140,70)[t]{$1$}
\Text(140,20)[t]{$2$}

\Text(120,70)[t]{$3$}
\Text(120,20)[t]{$4$}
\Text(260,70)[t]{$3$}
\Text(260,20)[t]{$4$}

\end{picture}

\vspace*{3mm}
\begin{picture}(260,100)(0,0)
    \SetWidth{1.0}
    \SetColor{Black}

\Text(0,50)[t]{$D$}

\Line(10,20)(110,20)
\Line(10,70)(110,70)
\Gluon(60,20)(60,70){4}{6}
\DashLine(35,20)(70,0){3}

\Line(150,20)(250,20)
\Line(150,70)(250,70)
\Gluon(200,20)(200,70){4}{5}
\DashLine(225,20)(260,0){3}

\Text(0,70)[t]{$1$}
\Text(0,20)[t]{$2$}
\Text(140,70)[t]{$1$}
\Text(140,20)[t]{$2$}

\Text(120,70)[t]{$3$}
\Text(120,20)[t]{$4$}
\Text(260,70)[t]{$3$}
\Text(260,20)[t]{$4$}

\end{picture}

\caption[]{$t$ channel gluon exchange graphs contributing to the subprocesses $q_i+\bar q_j
\rightarrow \phi +q_k+\bar q_l$, $q_i + q_j \rightarrow \phi +q_k + q_l$ and
$\overline q_i + \overline q_j \rightarrow \phi +\overline q_k + \overline q_l$. }
\label{fig9}
\end{center}
\end{figure}
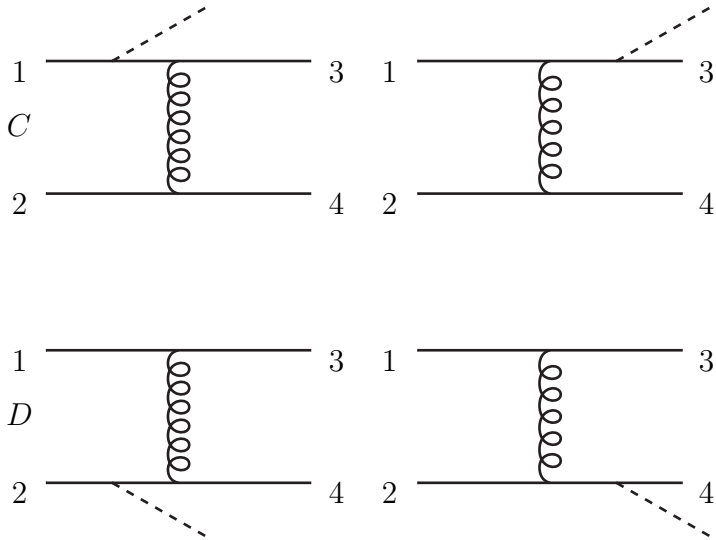


\begin{figure}
\begin{center}
 \begin{picture}(260,85)(0,0)
    \SetWidth{1.0}
    \SetColor{Black}

\Text(0,50)[t]{$E$}

\Line(10,20)(110,20)
\Line(10,70)(110,70)
\Gluon(60,20)(60,70){4}{6}
\DashLine(35,70)(70,90){3}

\Line(150,20)(250,20)
\Line(150,70)(250,70)
\Gluon(200,20)(200,70){4}{5}
\DashLine(225,70)(260,90){3}

\Text(0,70)[t]{$1$}
\Text(0,20)[t]{$2$}
\Text(140,70)[t]{$1$}
\Text(140,20)[t]{$2$}

\Text(120,70)[t]{$4$}
\Text(120,20)[t]{$3$}
\Text(260,70)[t]{$4$}
\Text(260,20)[t]{$3$}

\end{picture}

\vspace*{3mm}
\begin{picture}(260,100)(0,0)
    \SetWidth{1.0}
    \SetColor{Black}

\Text(0,50)[t]{$F$}

\Line(10,20)(110,20)
\Line(10,70)(110,70)
\Gluon(60,20)(60,70){4}{6}
\DashLine(35,20)(70,0){3}

\Line(150,20)(250,20)
\Line(150,70)(250,70)
\Gluon(200,20)(200,70){4}{5}
\DashLine(225,20)(260,0){3}

\Text(0,70)[t]{$1$}
\Text(0,20)[t]{$2$}
\Text(140,70)[t]{$1$}
\Text(140,20)[t]{$2$}

\Text(120,70)[t]{$4$}
\Text(120,20)[t]{$3$}
\Text(260,70)[t]{$4$}
\Text(260,20)[t]{$3$}

\end{picture}
\caption[]{$t$ channel gluon exchange graphs contributing to the subprocesses
$q(\overline q) + q (\overline q)\rightarrow \phi +q(\overline q) + q(\overline q)$ with identical quarks (anti-quarks). }
\label{fig10}
\end{center}
\end{figure}
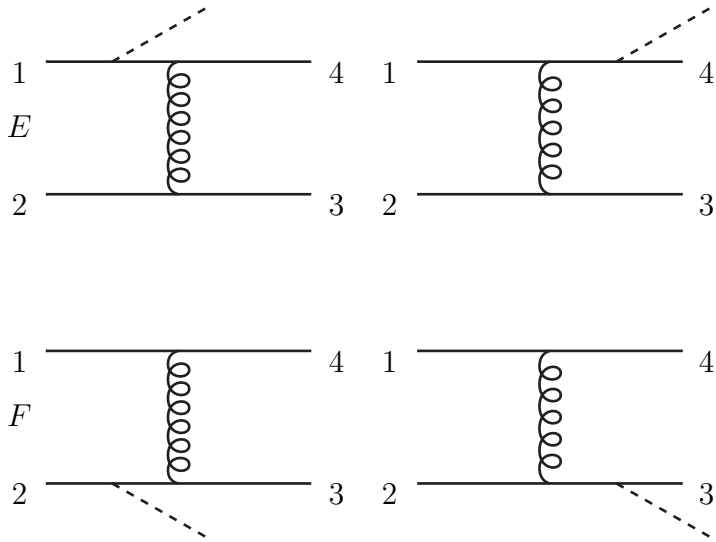

\begin{figure}
\vspace{-3cm}
\centering
                     \includegraphics[scale=0.7]{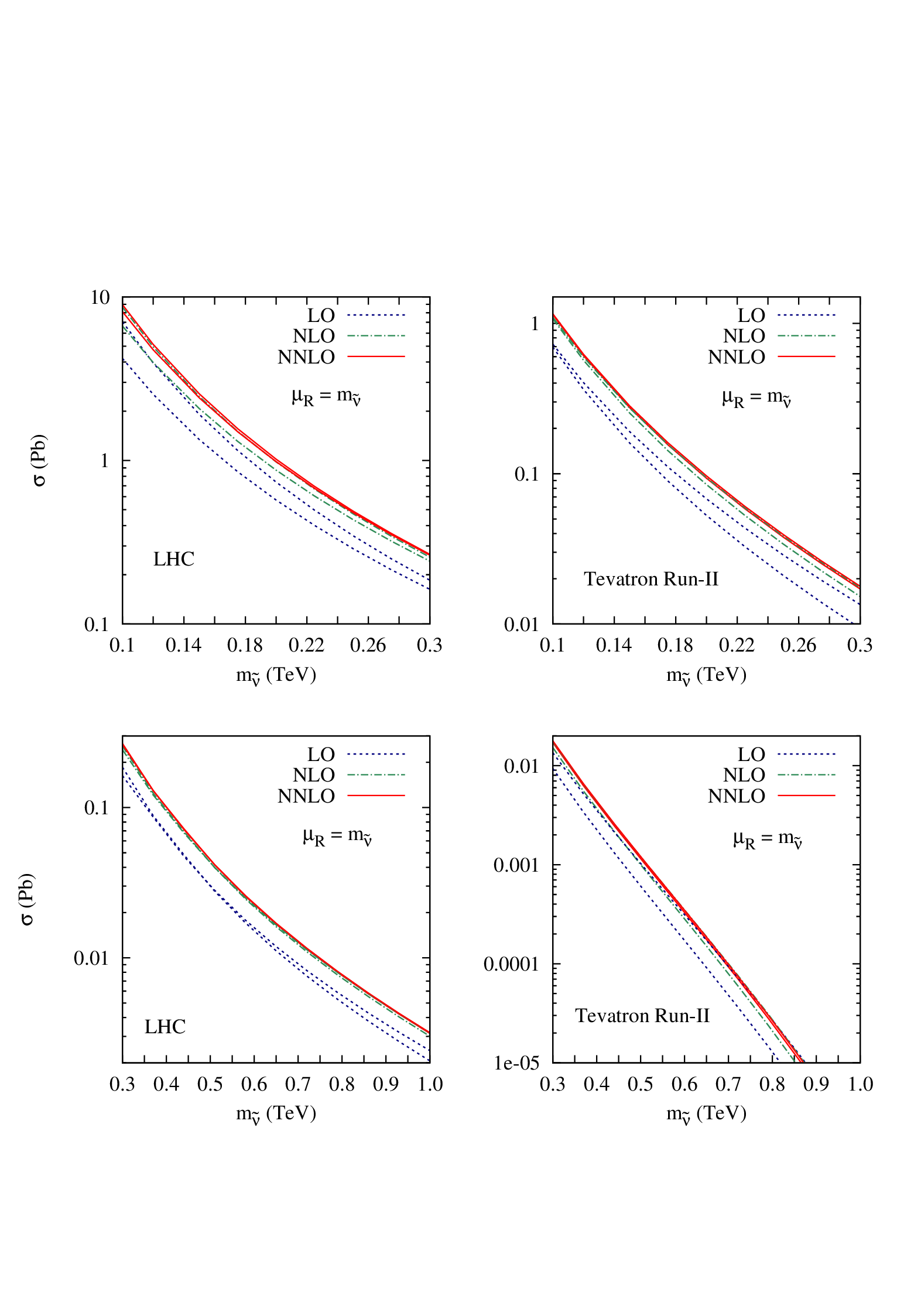}
\vspace{-4cm}
\caption{\em Total cross-section for the ${\tilde \nu}$ production.
              The upper (lower) set of lines correspond to the
factorisation scale $\mu_F = m_{\tilde{\nu}}(0.1 m_{\tilde{\nu}})$.
 }
\label{fig:l0}
\end{figure}
\begin{figure}
\vspace{-3cm}
                     \includegraphics[scale=0.7]{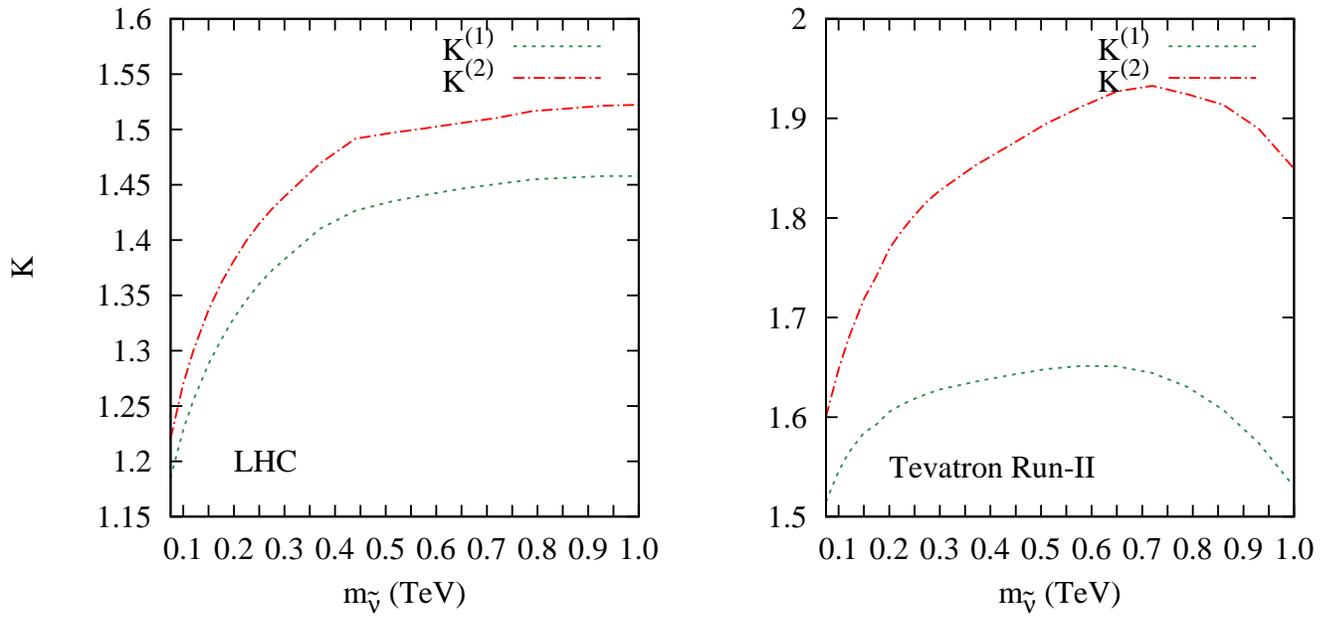}
\vspace{-13cm}
\caption{\em K-factor for the ${\tilde \nu}$ production.
 }
\label{fig:kl0}
\end{figure}
\begin{figure}
\vspace{-3cm}
                     \includegraphics[scale=0.7]{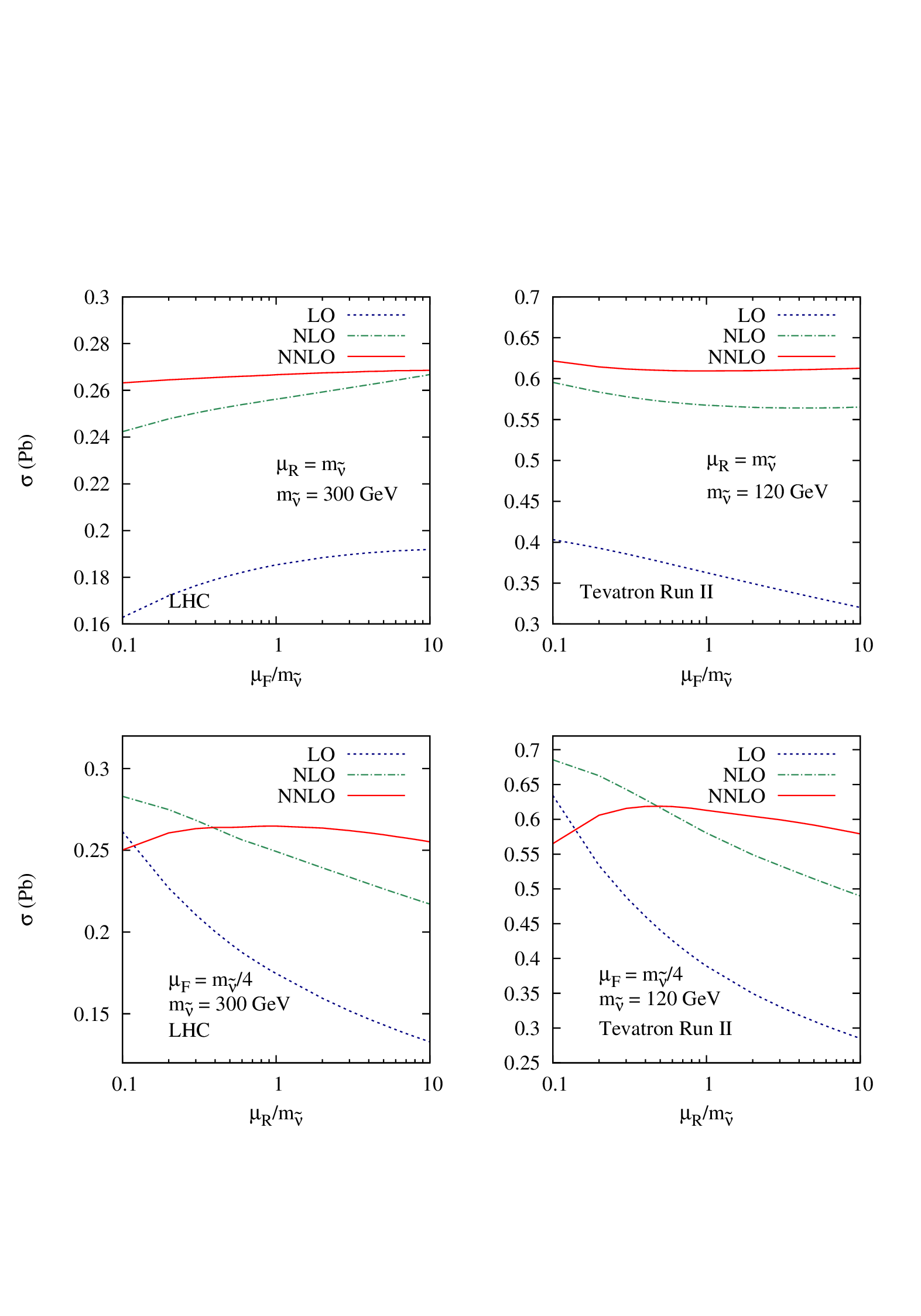}
\vspace{-4cm}
\caption{\em $\mu_F$ and $\mu_R$ variations for the ${\tilde \nu}$ production.
 }
\label{fig:ml0}
\end{figure}
\begin{figure}
\vspace{-3cm}
                     \includegraphics[scale=0.7]{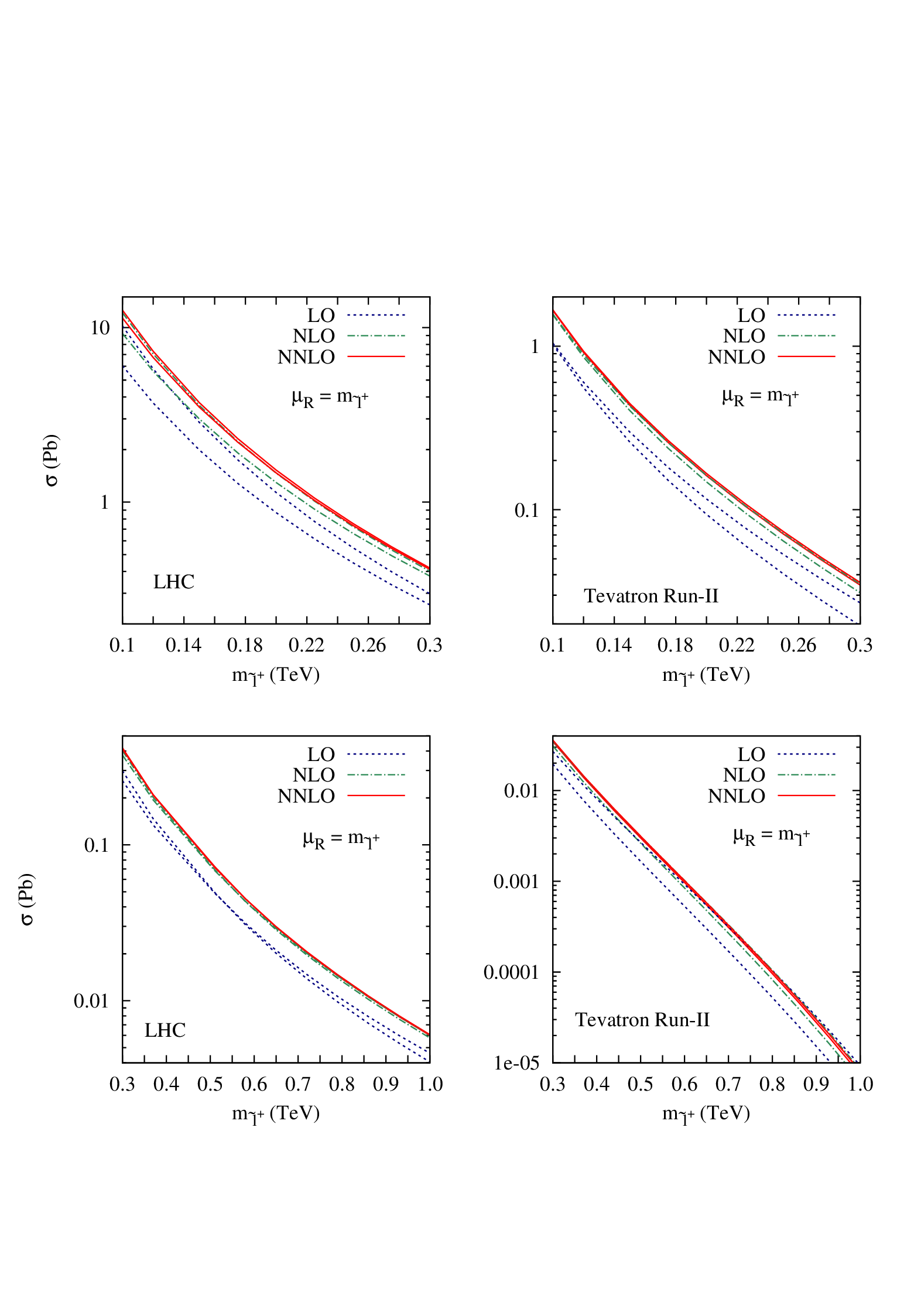}
\vspace{-4cm}
\caption{\em Total cross-section for the ${\tilde{\ell}^+}$ production.
              The upper (lower) set of lines correspond to the
factorisation scale $\mu_F = m_{\tilde{\ell}^+}(0.1 m_{\tilde{\nu}})$.
 }
\label{fig:lp}
\end{figure}
\begin{figure}
\vspace{-3cm}
                     \includegraphics[scale=0.7]{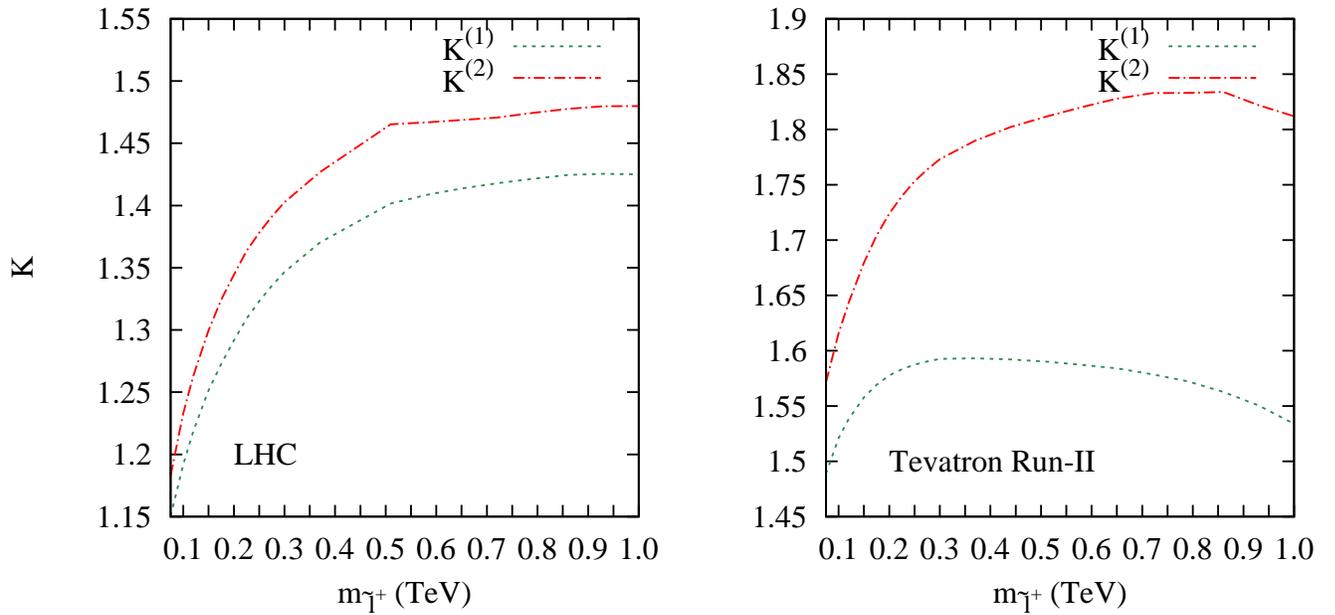}
\vspace{-13cm}
\caption{\em K-factor for the ${\tilde{\ell}^+}$ production.
 }
\label{fig:klp}
\end{figure}
\begin{figure}
\vspace{-3cm}
                     \includegraphics[scale=0.7]{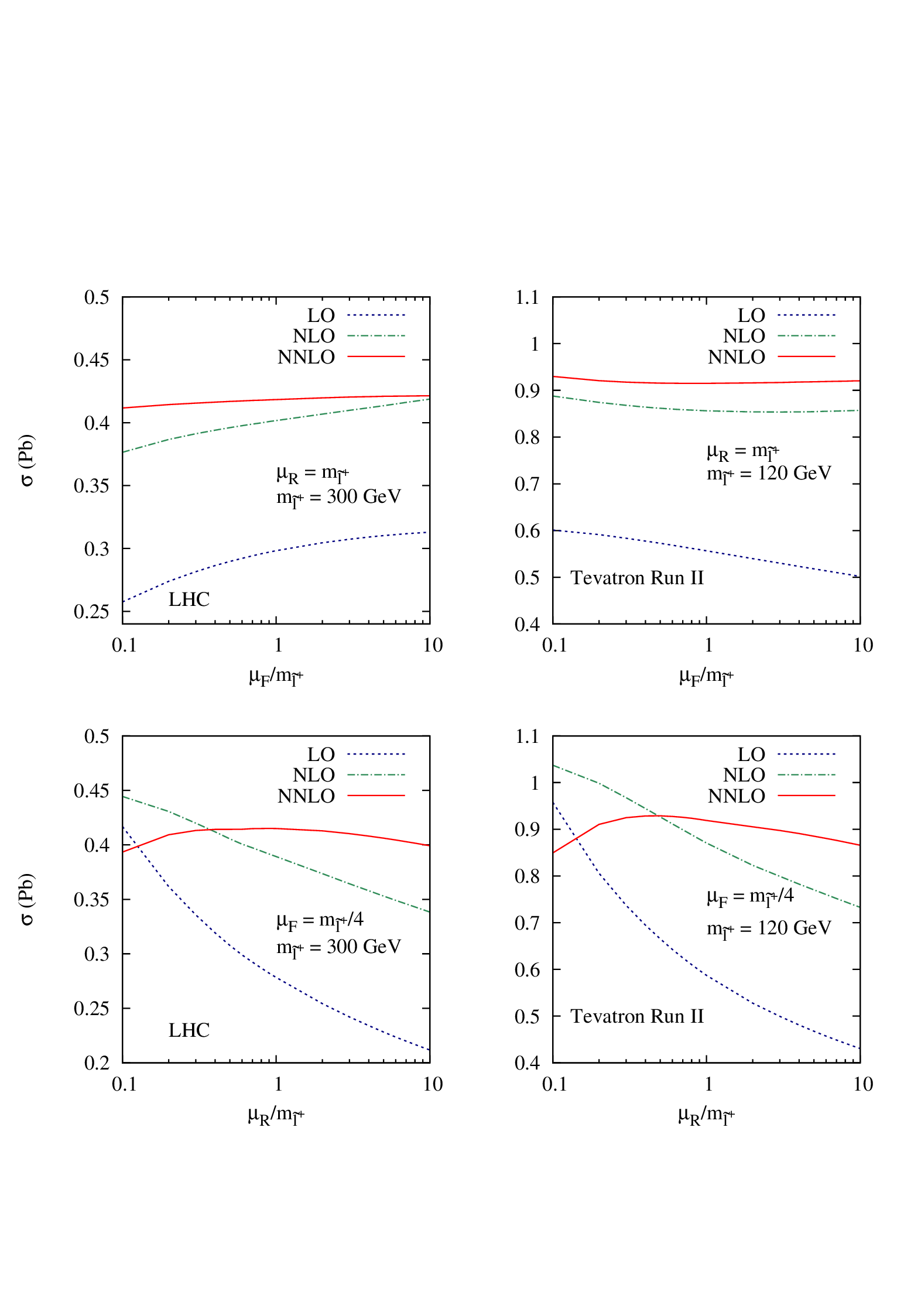}
\vspace{-4cm}
\caption{\em $\mu_F$ and $\mu_R$ variations for the ${\tilde{\ell}^+}$ production.
 }
\label{fig:mlp}
\end{figure}
\begin{figure}
\vspace{-3cm}
                     \includegraphics[scale=0.7]{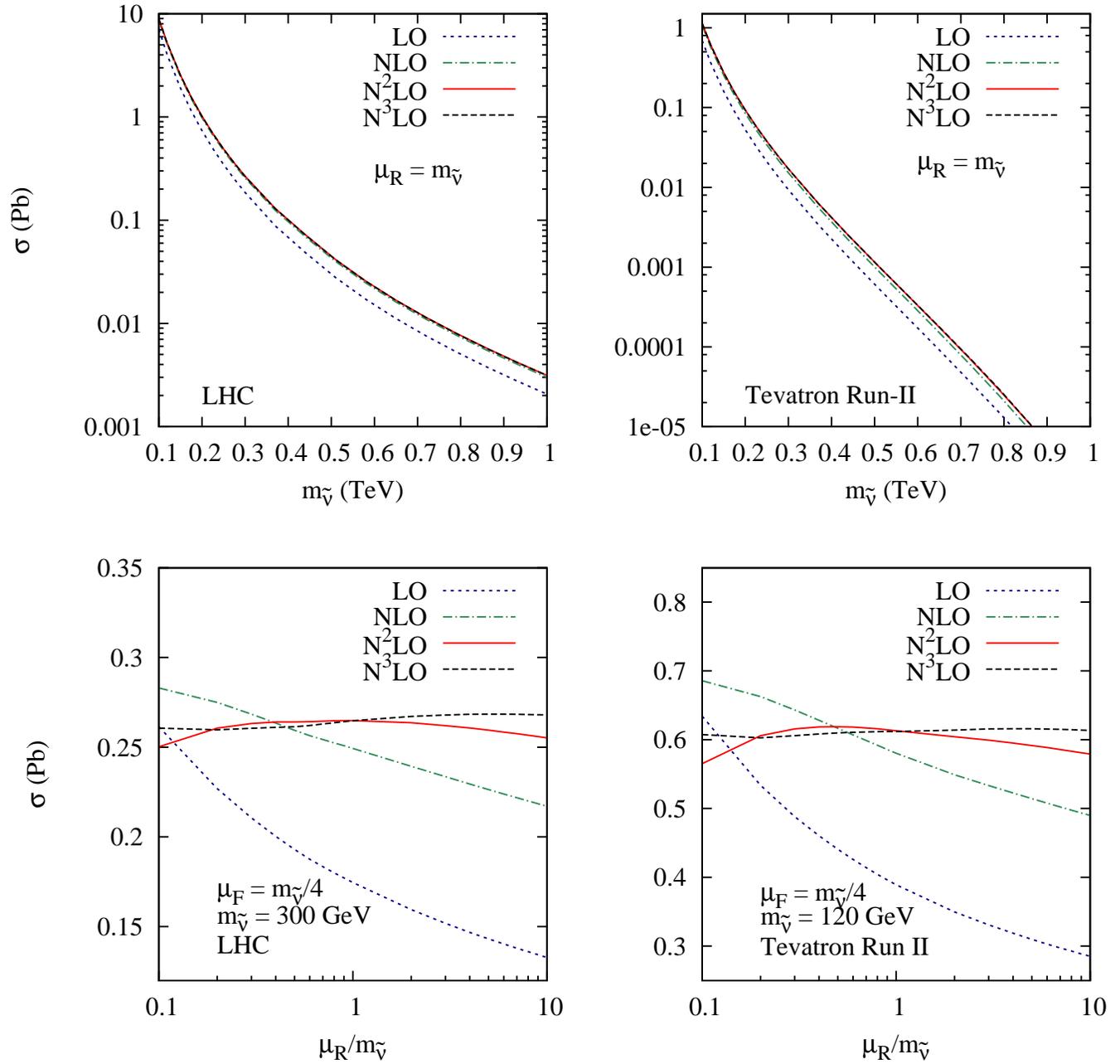}
\vspace{-4cm}
\caption{\em Total cross section and $\mu_R$ variations for the ${\tilde{\nu}}$ production.
 }
\label{fig:n3l0}
\end{figure}

\end{document}